\documentclass[a4paper,fleqn]{cas-sc}

\usepackage[numbers]{natbib}
\usepackage[utf8]{inputenc} 
\usepackage{graphicx}
\usepackage{caption}
\usepackage{subcaption}
\usepackage{bm}
\usepackage{physics}
\usepackage{amsmath}
\usepackage{mathtools}
\usepackage[svgnames]{xcolor}
\usepackage{amssymb}
\usepackage{amsthm}
\usepackage{siunitx}
\usepackage{microtype}      
\usepackage{amsfonts}       
\usepackage{nicefrac}       
\usepackage{float}
\usepackage[section]{placeins}

\newcommand{\bbA}{\mathbb{A}}

\newcommand{\bbT}{\mathbb{T}}
\newcommand{\bbS}{\mathbb{S}}
\newcommand{\bbL}{\mathbb{L}}
\newcommand{\bbI}{\mathbb{I}}

\newcommand{\bbK}{\mathbb{K}}
\newcommand{\bbH}{\mathbb{H}}
\newcommand{\bbB}{\mathbb{B}}
\newcommand{\xx}{\bm{x}}

\newcommand{\vv}{\bm{v}}
\newcommand{\ff}{\bm{f}}

\newcommand{\calU}{\mathcal{U}}

\newcommand{\calP}{\mathcal{P}}

\newtheorem{state}{Statement}

\DeclareMathOperator{\dev}{dev}

\begin{document}
	\let\WriteBookmarks\relax
	\def\floatpagepagefraction{1}
	\def\textpagefraction{.001}

	\hypersetup{
		pdftitle={Unified description of fluids and solids in Smoothed Particle Hydrodynamics}
		pdfsubject={}
	pdfauthor={O. Kincl, I. Peshkov, M. Pavelka, and V. Klika},
	pdfkeywords={SPH; SHTC; solid mechanics; fluid dynamics; energy-conservation; tensile instability}
}

\shorttitle{SHTC-SPH}

\shortauthors{O. Kincl, I. Peshkov, M. Pavelka, and V. Klika}

\title [mode = title]{Unified description of fluids and solids in Smoothed Particle Hydrodynamics}



%

\author[1]{Ond{\v r}ej Kincl}

\cormark[1]


\ead{ondrej.kincl.6@gmail.com}


\affiliation[1]{organization={Mathematical Institute, Faculty of Mathematics, Charles University},
	addressline={Sokolovsk\'{a} 83}, 
	city={Prague},
	postcode={186 75}, 
	country={Czech Republic}}

\cortext[1]{Corresponding author}

\author[2]{Ilya Peshkov}[
	orcid=0000-0001-8285-0639
	]



\ead{ilya.peshkov@unitn.it}


\affiliation[2]{organization={University of Trento},
	addressline={Department of Civil, Environmental and Mechanical Engineering, Via Mesiano 77}, 
	city={Trento},
	postcode={38123}, 
	country={Italy}}

\author[1]{Michal Pavelka}[
	orcid=0000-0003-0605-6737
	]



\ead{pavelka@karlin.mff.cuni.cz}


\author[3]{V{\'a}clav Klika}



\ead{vaclav.klika@cvut.cz}


\affiliation[3]{organization={Czech Technical University in Prague},
	addressline={Dept. of Mathematics, FNSPE, Trojanova 13}, 
	city={Prague},
	postcode={120 00}, 
	country={Czech Republic}}



\begin{abstract}
	Smoothed Particle Hydrodynamics (SPH) methods are advantageous in simulations of fluids in domains with free boundary. Special SPH methods have also been developed to simulate solids. However, there are situations where the matter behaves partly as a fluid and partly as a solid, for instance, the solidification front in 3D printing, or any system involving both fluid and solid phases. We develop an SPH-like method that is suitable for both fluids and solids at the same time. Instead of the typical discretization of hydrodynamics, we discretize the Symmetric Hyperbolic Thermodynamically Compatible equations (SHTC), which describe both fluids, elastic solids, and visco-elasto-plastic solids within a single framework. The resulting SHTC-SPH method is then tested on various benchmarks from the hydrodynamics and dynamics of solids and shows remarkable agreement with the data.
\end{abstract}




\maketitle


\tableofcontents

\section{Introduction}

We continue investigating different numerical strategies for the discretization of the unified formulation of continuum fluid and solid mechanics \cite{HPR2016,DPRZ2016}, which can describe flows of Newtonian and non-Newtonian fluids \cite{nonNewtonian2021}, as well as deformations of elastoplastic solids \cite{Hyper-Hypo2019} in a single system of first-order hyperbolic partial differential equations. In this paper, we are particularly interested in the capabilities of the Smoothed Particle Hydrodynamics approach to capture the solution to the unified model in both fluid and solid regimes. Because the non-dissipative part of the model (all differential terms) belongs to the class of Symmetric Hyperbolic Thermodynamically compatible (SHTC) equations \cite{SHTC-GENERIC-CMAT,Rom1998,GodRom-elements,GodRom1995,God1961}, we shall also refer to the unified model as the SHTC equations. Moreover, as shown in \cite{SHTC-GENERIC-CMAT} the SHTC equations can also be seen as a particular realization of the GENERIC (General Equation for Non-Equilibrium Reversible-Irreversible Coupling) approach \cite{go,og,pkg} to non-equilibrium thermodynamics. In the long-term perspective, we, therefore, are interested in developing an SPH Hamiltonian integrator that respects various properties of the continuous equations (differential constraints, Jacobi identity, etc.) at the discrete level. This goal is partially addressed in this paper. 

Previously, the unified model of continuum mechanics was discretized using various \emph{mesh-based} techniques including Godunov-type finite volume methods and Discontinuous Galerkin methods \cite{DPRZ2016}, Arbitrary Lagrangian Eulerian methods \cite{Hyper-Hypo2019,Busto2020}, a finite volume methods in the Updated Lagrangian formulation with a high-order IMEX time integrator \cite{LGPR2022}, semi-implicit staggered finite volume method \cite{SIGPR2021} for low-Mach problems, thermodynamically compatible finite volume scheme \cite{HTC2022}. 

This time, we turn to the discretization of the SHTC equations with \emph{mesh-free} methods and, in particular, with the Smoothed Particle Hydrodynamics (SPH), which is a particle-based numerical method for partial differential equations introduced by Gingold and Monaghan in 1977 \cite{sph}. The method allows for an elegant treatment of complex time-dependent geometries. This feature makes it attractive for problems involving fluid-structure interactions \cite{antoci2007numerical} and multiphase flows \cite{monaghan1995sph}. Despite its name, the method is also applicable to solids \cite{Gil2016}. We refer, for example, to \cite{monaghan2005} for a comprehensive review. 

Although SPH methods were successfully applied to simulate fluids and solids, the schemes and equations were rather different. For example, fluid mechanics equations are formulated in the Eulerian frame, while solid mechanics equations are traditionally formulated in the Lagrangian frame. Our ultimate goal, therefore, to develop a single reliable scheme that works in both fluid and solid regimes of the SHTC equations, is far from being trivial. For example, such a goal is very appealing from the perspective of modeling material flows that exhibit coexistence of the fluid and solid states, as well as mutual transformations, e.g. selective laser metal printing (3D printing of metals), flows of viscoplastic fluids, granular flows \cite{Forterre2013}, landslides and avalanches, ice formation, etc. 

Despite the fact that the SHTC equations are formulated in the Eulerian frame (which is necessary for a fluid-like motion), these equations also have a particle-like, and therefore Lagrangian, nature, which was discussed in \cite{HYP2016}. In particular, the main field of SHTC equations that makes it possible to describe fluids and solids at once is the distortion field $ \bbA$. Distortion can be seen as a field of infinitesimal local basis triads, which are allowed to arbitrary rearrange with their neighbors and thus exhibit the particle-like nature. Such a continuum description of matter was in particular inspired by Frenkel's idea to characterize the fluidity 
of the liquids by the so-called characteristic particle rearrangement time $ \tau $ \cite{Frenkel1955,brazhkin2012two}. Moreover, the SHTC equations can be derived by the transformation of the Lagrangian Hamiltonian continuum mechanics to the Eulerian continuum \cite{LagEul}. 
Due to such a particle-like nature of the SHTC equations and the freedom in the rearrangements of the infinitesimal material elements, the mesh-based methods can not completely capture the microscopic dynamics of the SHTC equations, but it can only be revealed with a particle-based method, e.g. SPH.  

The numerical scheme proposed in this paper is based on an explicit variant of SPH. First, we perform the spatial semidiscretization of the reversible part of the SHTC system. The exact form of discrete operators is derived from a potential, which guarantees the conservation of energy. We obtain a system of ordinary differential equations, for which we find an efficient time-reversible integrator. We also discuss the problem of tensile instability and suggest a solution without interfering with the conservative properties of SPH. Finally, the irreversible part is added using the Runga-Kutta scheme as time integrator. The last section is devoted to the validity tests.

\section{Governing PDEs}

The unified model of continuum fluid and solid mechanics is formulated in the Eulerian frame in a Cartesian coordinate system $ \bm{x} = \{x_1,x_2,x_3\} $ as follows \cite{HPR2016,DPRZ2016}
\begin{subequations}\label{eqn.PDE}
	\begin{align}
		\frac{\partial \rho}{\partial t} &+ \nabla \cdot (\rho \vv) = 0 ,\\[2mm]
		\frac{\partial (\rho \vv)}{\partial t} &+ \nabla \cdot (\vv \otimes \rho \vv - \bbS) 
		= 0,\\[2mm]
		\frac{\partial \bbA}{\partial t} &+ \vv \cdot \nabla \bbA + \bbA \nabla \vv = 
		-\frac{1}{\theta} E_{\bbA},\label{eqn.PDE.A}\\[2mm]
		\frac{\partial (\rho E)}{\partial t} &+ \nabla \cdot (\vv \rho E - \bbS \, \vv) 
		= 0,
	\end{align}
\end{subequations}
where $ \rho $ is the mass density of the material, $ \vv = \{v_1,v_2,v_3\} $ is the velocity 
field, $ \vv \otimes\rho\vv = \{\rho v^i v^j\} $, $ \bbA = \{A^{ij}\}$ is the distortion field, $ 
\nabla\vv = \{\frac{\partial v^i}{\partial x^j}\} $, $ \vv\cdot\nabla = v^i 
\frac{\partial}{\partial x^i} $, $ \bbS = \{ S^{ij}\}$ is the total stress tensor whose 
specification depends on the material under consideration and is defined by the total energy 
density specification $ E = E(\rho,\vv,\bbA) = \epsilon(\rho,\bbA) + \frac12 \Vert \vv \Vert^2$, 
see Section\,\ref{sec.constitut}. In addition, $ \epsilon(\rho,\bbA) $ is the internal energy that 
must be specified by the user. 

The left-hand side of the equations is the reversible part of the time evolution and can be derived either from the variational principle or can be generated by Poisson brackets \cite{SHTC-GENERIC-CMAT}. This part describes the elasticity of the material. The right-hand side is characterized by the relaxation term in the distortion equation. Here, $ E_{\bbA} = \frac{\partial E}{\partial \bbA}  $ and is essentially the Lagrangian stress tensor (first Piola-Kirchhoff stress), while the scalar $ \theta = \theta(\rho,\tau,\bbA) \geq 0 $ is a 
relaxation function which depends on the state variables and some material constants. In 
particular, $ \theta \sim \tau $, where $ \tau $ is the strain relaxation time and one of the key 
elements of the SHTC model to describe fluids and solids. For example, in this framework, fluids 
can be seen as the relaxation limit (small relaxation time $ 0 < \tau \ll \infty $) of a 
solid when the shear stresses are strongly relaxed (``melted'' solid). For Newtonian fluids $ \tau 
$ can be taken constant, while for non-Newtonian fluids and elastoplastic solids $ \tau $ is the 
function of the stress state \cite{nonNewtonian2021,Hyper-Hypo2019}, and therefore of the 
distortion field $ \tau = \tau(\bbA) $ (as well as other parameters, e.g. temperature). For 
simplicity, we ignore the heat transfer effect which is also described by hyperbolic relaxation 
equations in the SHTC framework \cite{SIGPR2021,SHTC-GENERIC-CMAT,pkg} ,as well as the materials 
are considered as isothermal. The heat conduction will be 
included in a follow up paper.

The following section contains a numerical method (SHTC-SPH) that finds approximate solutions of 
the SHTC equations \eqref{eqn.PDE} in both the fluid and solid regimes.

\section{The SHTC-SPH Method}
In order to address the SHTC equations \eqref{eqn.PDE}, which contain the distortion field (unlike hydrodynamics), we have to define a discrete analogy of the continuous distortion. But before that, let us first recall the standard construction of SPH via smoothing kernels. 

The SPH is based on \emph{smoothing kernels} to calculate the influence of a particle on its surroundings. In this paper, we will use Wendland's quintic kernel, which reads
\begin{equation}
	w(r) = \begin{cases}
		\frac{\alpha_d}{h^d}  \left(1 - \frac{r}{2h}\right)^4\left(1 + \frac{2r}{h} \right), & 
		r \leq 2h\\
		0, & r \geq 2h
	\end{cases}
	\label{eq:wendland}
\end{equation}
where $r$ is the distance from the center of the particle, $h$ is the \textit{smoothing length} and $d$ is the dimension. The constant $\alpha_d$ has the following values:
\begin{equation*}
	\begin{cases}
		\alpha_2 = \frac{7}{4\pi}\\
		\alpha_3 = \frac{21}{16\pi}
	\end{cases}.
\end{equation*}
Following Violeau \cite{violeau}, we will use the notation 
\begin{equation*}
	w_{ab} = w(r_{ab}), \quad w'_{ab} = \derivative{w}{r} \left(r_{ab}\right)
\end{equation*}
where $r_{ab} = |\bm{x}_a - \bm{x}_b| = |\bm{x}_{ab}|$ is the distance between two particles with positions $\bm{x}_a, \bm{x}_b$ in the Eulerian frame. Furthermore, let us also denote
\begin{equation*}
	\nabla w_{ab} = w'_{ab} \frac{\xx_{ab}}{r_{ab}}. 
\end{equation*}
Realizing that
\begin{equation}
	\frac{w'(r)}{r} = \begin{cases}
		-\frac{10\alpha_d}{h^{d+1}}  \left(1 - \frac{r}{2h}\right)^3, & r \leq 2h\\
		0, & r \geq 2h
	\end{cases}
	\label{eq:rDwendland}
\end{equation}
we can implement $\nabla w_{ab}$ in a way that avoids potential division by $r = 0$.

Moreover, in the initial state, the particles are placed in a regular pattern, filling a domain $\Omega_0$ such that every particle occupies a volume $V_0 = \delta r^d$, where $\delta r > 0$ is the spatial step of the simulation. In the 2D case, an isometric grid arrangement is used in this paper, while in the 3D case, the particles are initially distributed in a body-centered cubic crystal. We can now proceed to the definition of the discrete state variables.

\subsection{Discrete density}
Mass density can be approximated, using interpolation by smoothing kernel density $\rho_a$ at $\xx_a$, as
\begin{equation}
	\rho_{a} = \sum_b m_b w_{ab} + C_{\rho, a}
	\label{eq:rho}
\end{equation}
where $C_{\rho, a}$ is a time-independent parameter that enforces the equality of $\rho_a$ with the 
reference density $\rho_0$ at the initial time instant $ t=0 $ (this is necessary to obtain the 
vanishing internal energy of the free 
surface particles). We assume that the masses of the particles $m_b$ are positive and do not depend 
on time \cite{sph-reversible}. In all the examples presented in this paper, we use $m_b = \rho_0 
V_0\big|_b$ where the volume of the particles follows from $V_a=(\sum_b w_{ab})^{-1}$. 

A straightforward computation then yields the following formula for the differential of the discrete density.
\begin{state}
	Assuming $r_{ab} > 0$ for each pair of particles, the total differential of $\rho_{a}$ according to formula \eqref{eq:rho} with respect to the positions of the particles $\xx$ is 
	\begin{equation}
		\dd{\rho}_a = \sum_b m_b  \dd{\xx}_{ab} \cdot \nabla w_{ab}
		\label{eq:drho} 
	\end{equation}
\end{state}
\subsection{Discrete distortion}
An important variable in the SHTC equations is the \textit{distortion} matrix. In the reversible (elastic) case, when it can be thought of as the inverse of the deformation tensor, it satisfies the following equation \cite[Chap 3]{pkg}
\begin{equation}
	\dot{\bbA} = - \bbA \bbL,
	\label{eq:A_dot}
\end{equation}
where $\bbL$ is the velocity gradient and the dot denotes the material time derivative. Using a renormalized SPH gradient \cite{bonet1999variational, vila1999particle} inspired by a work by Falk and Langer \cite{falk-langer}, we can approximate this quantity as
\begin{equation}
	\bbL_a = \left( \sum_b m_b \vv_{ab} \otimes \nabla w_{ab}\right) \left( \sum_b m_b \xx_{ab} \otimes \nabla w_{ab} \right)^{-1}.
	\label{eq:L}
\end{equation}
The following statement explains that $\bbL_a$ can be found by approximating a differential formula $\dd{\vv} = \bbL \dd{\xx}$:
\begin{state} \label{state:A}
	Assume $r_{ab} > 0$ for each pair of particles and for any fixed particle $a$ that there are $d$ linearly independent vectors $\xx_{ab}$ satisfying $r_{ab} < 2h$. (In other words, particle $a$ and its neighbors must not be co-planar in 3D or co-linear in 2D.) Then the matrix inverse in \eqref{eq:L} exists and $\bbL_a$ is the unique solution of the overdetermined system
	\begin{equation*}
		\bbL_a \xx_{ab} \doteq \vv_{ab}
	\end{equation*}
	in the sense of weighted least squares with weights $m_b \frac{|w'_{ab}|}{r_{ab}}$.
\end{state} 
\begin{proof}
	Note that $|w'_{ab}| = -w'_{ab}$. It is clear from the assumptions that the matrix \begin{equation*}
		-\sum_b m_b \xx_{ab} \otimes \nabla w_{ab} = \sum_b \frac{m_b |w'_{ab}|}{r_{ab}} \left( \xx_{ab} \otimes \xx_{ab} \right)
	\end{equation*}    
	is positive definite and thus invertible. Now, the global minimimum of coercive and differentiable function
	\begin{equation*}
		\mathcal{E}(\bbL_a) = -\frac{1}{2}\sum_b m_b \frac{w'_{ab}}{r_{ab}} |\bbL_a \xx_{ab} - \vv_{ab}|^2
	\end{equation*}
	with respect to $\bbL_a$ exists and satisfies
	\begin{equation*}
		\begin{split}
			0 &= \dd{\mathcal{E}_a} = -\sum_b m_b \frac{w'_{ab}}{r_{ab}} (\bbL_a \xx_{ab} - \vv_{ab}) \cdot  \dd{\bbL_a} \xx_{ab}\\
			&= -\dd{\bbL_a} : \sum_b m_b (\bbL_a \xx_{ab} - \vv_{ab}) \otimes \nabla w_{ab}, \qquad \forall \dd{\bbL_a}.
		\end{split}
	\end{equation*}
	This immediately yields \eqref{eq:L}.
\end{proof}
As a corollary of Statement \ref{state:A}, it is clear that the definition \eqref{eq:L} is first-order exact. In fact, $\bbL_a \xx_{ab} = \vv_{ab}$ will be solved exactly by least squares, provided that there is an exact solution, which occurs exactly when $\vv$ can be written as a linear function of $\xx$. 

Combining \eqref{eq:A_dot} with \eqref{eq:L}, we obtain the evolution of $\bbA_a$ in the form
\begin{equation}
	\dot{\bbA}_a = - \bbA_a \left( \sum_b m_b \vv_{ab} \otimes \nabla w_{ab}\right) \left( \sum_b m_b \xx_{ab} \otimes \nabla w_{ab} \right)^{-1}.
	\label{eq:A_dot_a}
\end{equation}
We can rephrase this using a linear form that relates changes of $\bbA_a$ to variations of $\xx_a$:
\begin{equation}
	\dd {\bbA}_a = -\bbA_a \left(\sum_b m_b \dd{\xx}_{ab} \otimes \nabla w_{ab} \right) \left( \sum_b m_b \xx_{ab} \otimes \nabla w_{ab} \right)^{-1},
	\label{eq:dA}
\end{equation}
which defines $\bbA_a$ provided that its initial value is specified and the particle trajectories are known. Here, we have a subtle problem because we do not have any guarantee that the right-hand side in \eqref{eq:dA} is integrable. Therefore, when $\Gamma$ is a closed loop in the configuration space, we, in general, have the following
\begin{equation*}
	\int_\Gamma  \dd {\bbA}_a \neq 0.
\end{equation*}
In other words, if \eqref{eq:dA} is used, the numerical distortion may not recover when the shape 
of a material does, potentially introducing some artificial irreversibility. Without resorting to 
the Lagrangian description, we have not found a satisfactory solution to this problem in pure 
Eulerian settings.

On a side note, instead of treating the density as a separate variable, it is possible to use 
\begin{equation*}
	\rho_a = \rho_0 \det \bbA_{a}. 
\end{equation*}
However, in our numerical experience, using formula \eqref{eq:rho} is more reliable.

\subsection{Reversible part of the SHTC-SPH equations}
Let us now consider the case of an elastic solid with internal energy
\begin{equation*}
	\calU = \int_{\Omega} \rho \; \epsilon  \dd{x}.
\end{equation*}
which we discretize as
\begin{equation*}
	\calU_h = \sum_a  m_a \epsilon_a,
\end{equation*}
where $\epsilon_a = \epsilon(\rho_a, \bbA_a)$ is the specific internal energy. For simplicity of 
notation, let us write
\begin{equation}
	\begin{split}
		\bbH_a &= \sum_b m_b \xx_{ab} \otimes \nabla w_{ab}\\
		\bbT_a &= -\epsilon_{\rho_a} \bbI + \bbA_{a}^T \epsilon_{\bbA_a}\bbH_a^{-1},
	\end{split}
	\label{eq:HT}
\end{equation}
where we employed the usual notation for partial derivatives $\epsilon_{\rho_a}=\frac{\partial \epsilon}{\partial \rho_a}$, $\epsilon_{\bbA_a}=\frac{\partial \epsilon}{\partial \bbA_a}$.  Now, using \eqref{eq:drho}, \eqref{eq:dA}, we find how $\calU$ varies when $\xx$ changes: 
\begin{equation}
	\begin{split}
		\dd \calU_h &= \sum_a  m_a \epsilon_{\rho_a} \dd{\rho_a} + \sum_{a} m_a \epsilon_{\bbA_a} : \dd \bbA_a \\
		&= \sum_{a,b} m_a m_b \bigg( \epsilon_{\rho_a} \nabla w_{ab} \cdot \dd{\xx}_{ab}
		- \epsilon_{\bbA_{a}} : \bbA_{a} \left(\dd{\xx}_{ab} \otimes \nabla w_{ab} \right)\bbH_a^{-1} \bigg)\\
		&= -\sum_{a,b} m_a m_b \bbT_a \nabla w_{ab} \cdot \dd{\xx}_{ab}\\
		&= -\sum_{a,b} m_a m_b (\bbT_a + \bbT_b) \nabla w_{ab} \cdot \dd{\xx}_{a}.
	\end{split}
	\label{eq:dU}
\end{equation}
Combining \eqref{eq:A_dot_a}, \eqref{eq:dU} and $m_a \dot{\vv}_{a} = -\pdv{\calU_h}{\xx_a}$ (Newton's law), we obtain a system of ordinary differential equations:
\begin{equation}
	\begin{split}
		\dot{\xx}_{a} &= \vv_a\\
		\dot{\vv}_{a} &= \sum_{b} m_b (\bbT_a + \bbT_b) \nabla w_{ab} \\
		\dot{\bbA}_a &= -\bbA_a \bbL_a,
	\end{split}
	\label{eq:odes}
\end{equation}
which is the system of the SHTC-SPH ordinary differential equations approximating the SHTC equations.

One of the greatest assets of SHTC-SPH is that it enjoys various conservative properties.
\begin{state}
	The system of equations \eqref{eq:odes} satisfies the conservation of the
	\begin{itemize}
		\item energy
		$$ \mathcal{H}_h = \sum_{a} m_a \left( \frac{v_a^2}{2} + \epsilon_a\right),$$
		\item linear momentum
		$$  \bm{\mathcal{M}}_h = \sum_{a} m_a \vv_a ,$$
		\item angular momentum (provided that the shear stress tensor $\bbA_{a}^T \epsilon_{\bbA_{a}}$ is symmetric)
		$$ \bm{\mathcal{L}}_h = \sum_{a} m_a \xx_a \times \vv_a .$$
	\end{itemize}
\end{state}
\begin{proof}
	Conservation of energy follows from the construction, linear momentum is conserved due to antisymmetry $\nabla w_{ab} = -\nabla w_{ba}$. Likewise, showing conservation of angular momentum is easy because
	\begin{equation*}
		\begin{split}
			\dot{\mathcal{L}}^i_h &= \epsilon^{ijk}  \sum_{a} m_a x^j_a \dot{v}^k_a \\
			&= \epsilon^{ijk}  \sum_{a} m_a x^j_{a} \left(\sum_b m_b (T_a^{kl} + T_b^{kl}) \nabla_{ab}w^l\right) \\
			&= \epsilon^{ijk}  \sum_{a} m_a T_a^{kl}  \left(\sum_b m_b  x^i_{ab} \nabla_{ab}w^l\right)\\
			&= \epsilon^{ijk}  \sum_{a} m_a T_a^{kl}  H_a^{li}\\
			&= 0
		\end{split}
	\end{equation*}
	since the matrix $\bbT_a \bbH_a$ is symmetric.
\end{proof}
\subsection{Constitutive equations}\label{sec.constitut}
The set of equations \eqref{eq:odes} is incomplete until one specifies the dependence of internal energy $\epsilon$ on $\rho, \bbA$. In this paper, we assume the decomposition of energy into the bulk and shear components:
\begin{equation*}
	\epsilon(\rho, \bbA) = \epsilon_0(\rho) + \epsilon_s(\bbA).
\end{equation*}
For the bulk energy component, we can use a simple quadratic relation
\begin{equation}
	\epsilon_0(\rho) = \frac{c_0^2}{2}\left(\frac{\rho_0}{\rho} - 1\right)^2.
	\label{eq:eps0}
\end{equation}
where $c_0$ is the bulk speed of sound. This choice is not unique but serves well for the 
demonstration purposes. Alternatives exist (see, e.g. \cite{pkg}), but the distinction is typically 
less important for materials which are difficult to compress. From \eqref{eq:eps0}, we can deduce a 
formula for the pressure:
\begin{equation}
	p = \rho^2\pdv{\epsilon_0}{\rho} = c_0^2 \rho_0 \left(1 - \frac{\rho_0}{\rho}\right). \label{eq:p}
\end{equation}
For the shear elastic energy, we consider two variants. Following the papers \cite{DPRZ2016}, we 
can 
use
\begin{equation}
	\epsilon^{DPRZ}_s(\bbA) = \frac{c_s^2}{4}\| \dev(\bbA^T \bbA) \|^2_F,
	\label{eq:DPR}
\end{equation}
where $c_s$ is the shear speed of sound, $\| \cdot \|_F$ is the Frobenius norm and $$ \dev \mathbb{M} = \mathbb{M} - \frac{1}{3} (\trace{\mathbb{M}}) \; \bbI$$ denotes \textit{deviatoric part}. Plugging this into \eqref{eq:HT}, we get
\begin{equation}
	\bbT_a = -\frac{p_a}{\rho_a^2} \bbI + c_s^2 \bbA_a^T \bbA_a  \dev(\bbA_a^T \bbA_a) \bbH_a^{-1}.
\end{equation}
Although less convenient in the Eulerian setting, it is also possible to use the Neo-Hookean model
\begin{equation}
	\epsilon^{NH}_s(\bbA) = \frac{c_s^2}{2} \left( \trace \bbB_a - 3 + 2 \ln \det \bbA_{a} \right),
	\label{eq:NH}
\end{equation}
which yields:
\begin{equation}
	\bbT_a = -\frac{p_a}{\rho_a^2} \bbI - c_s^2 \left(\bbB_a - \bbI \right) \bbH_a^{-1},
\end{equation}
with $\bbB_a = \bbA_{a}^{-T} \bbA_{a}^{-1}$ being the left Cauchy-Green tensor and the pressure in 
a particle follows from \eqref{eq:p} as
\begin{equation*}
	p_a =c_0^2 \rho_0 \left(1 - \frac{\rho_0}{\rho_a}\right).
\end{equation*}

\subsection{Tensile penalty}
A common issue encountered in SPH is the \textit{tensile instability} --- a numerical artifact, which causes unwanted clumping of particles in regions of negative pressure. Due to non-linearities in formula \eqref{eq:rho}, density $\rho$ will increase slightly under tensile strain. This is usually not a problem for $p > 0$, however, for $p < 0$ particles can reduce their potential by tensile strain according to
$$ \pdv{\epsilon}{\rho} = \frac{p}{\rho^2}.$$
This often results in the formation of particle chains surrounded by void patches, which can 
eventually cause body tearing. There are a few remedies offered in the literature. Paper 
\cite{monaghan2000sph} recommends adding an artificial force that repels particles with abnormally 
small separation. Another possible treatment, called $\delta$-shifting \cite{sun2019consistent}, 
subjects the particles to artificial diffusion. In this paper, we suggest adding the following 
\textit{tensile penalty} term to the energy:
\begin{equation}
	\calP_h = \frac{1}{2}\sum_{a} m_a c_p^2 \left( \frac{\lambda_a}{\rho_0} \right)^2, 
	\label{eq:penalty}
\end{equation} 
where $c_p$ is a numerical parameter with velocity dimensions that determines the strength of 
anti-clumping forces. The variable $\lambda_a$ is defined by the relation
\begin{equation}
	\lambda_a = h\pdv{\rho_a}{h} + C_{\lambda, a} = \sum_{b} m_b h \pdv{w_{ab}}{h} + C_{\lambda, a}.
	\label{eq:lambda}
\end{equation}
Our idea is to describe clustering as a situation where $\rho_a$ increases when we take a smaller smoothing length $h$. By adding this energy term, we enforce a nice structure of the particles by keeping $\pdv{\rho_a}{h}$ small. Similarly to \eqref{eq:rho}, we add a time-independent parameter $C_{\lambda, a}$ to ensure that $\lambda_a = 0$ at the initial time. The potential $\calP_h$ then generates an additional force
\begin{equation*}
	-\pdv{\calP_h}{\xx_{a}} = - \sum_{b} m_a m_b c_p^2 \frac{\lambda_a + \lambda_b}{\rho_0^2} \, h \, \pdv{\nabla w_{ab}}{h}
\end{equation*}
so that the updated balance of momentum reads
\begin{equation}
	\begin{split}
		\dot{\vv}_{a} = &\sum_{b} m_b(\bbT_a  + \bbT_b) \nabla w_{ab} \\
		&- \sum_{b}  m_b c_p^2 \frac{\lambda_a + \lambda_b}{\rho_0^2} \, \frac{h}{r_{ab}} \frac{\partial^2 w_{ab}}{\partial r_{ab} \partial h} \xx_{ab}.
	\end{split}
	\label{eq:bom_with_tp}
\end{equation}
Restricting ourselves to the Wendland's kernel \eqref{eq:wendland}, we can provide these explicit formulas:
\begin{equation*}
	h\pdv{w}{h} = \begin{cases}
		\frac{7}{4\pi h^2}\left(1 - \frac{r}{2h}\right)^3\left(7 \left(\frac{r}{h}\right)^2 - \frac{3r}{h} - 2\right) & r \leq 2h\\
		0 & r \geq 2h
	\end{cases},
\end{equation*}
\begin{equation*}
	\frac{h}{r} \frac{\partial^2 w}{\partial r \partial h} = \begin{cases}
		-\frac{35}{4\pi h^4}\left(1 - \frac{r}{2h}\right)^2\left(\frac{7r}{h}- 8\right) & r \leq 2h\\
		0 & r \geq 2h
	\end{cases}
\end{equation*}
in 2D, and
\begin{equation*}
	h \pdv{w}{h} = \begin{cases}
		\frac{21}{32\pi h^3}\left(1 - \frac{r}{2h}\right)^3 \left(16 \left(\frac{r}{h}\right)^2 - \frac{9r}{h} - 6\right) & r \leq 2h\\
		0 & r \geq 2h
	\end{cases},
\end{equation*}
\begin{equation*}
	\frac{h}{r} \frac{\partial^2 w}{\partial r \partial h} = \begin{cases}
		-\frac{105}{16\pi h^5} \left(1 - \frac{r}{2h}\right)^2  \left(\frac{4r}{h} - 5\right)& r \leq 2h \\
		0 & r \geq 2h
	\end{cases}
\end{equation*}
in 3D. 

\begin{figure}[!hbt]
	\includegraphics[draft=false,width=0.49\columnwidth]{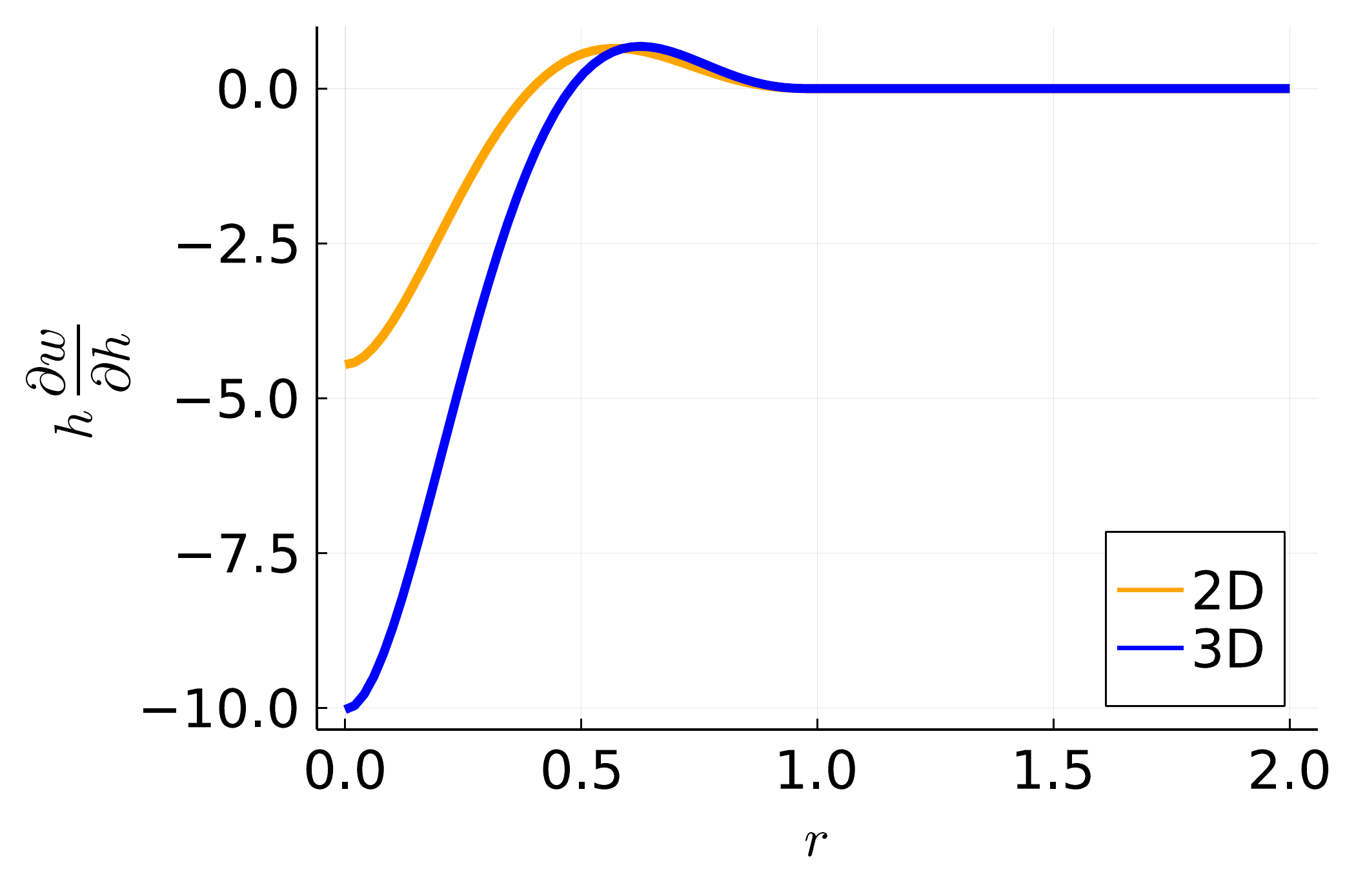}
	\includegraphics[draft=false,width=0.49\columnwidth]{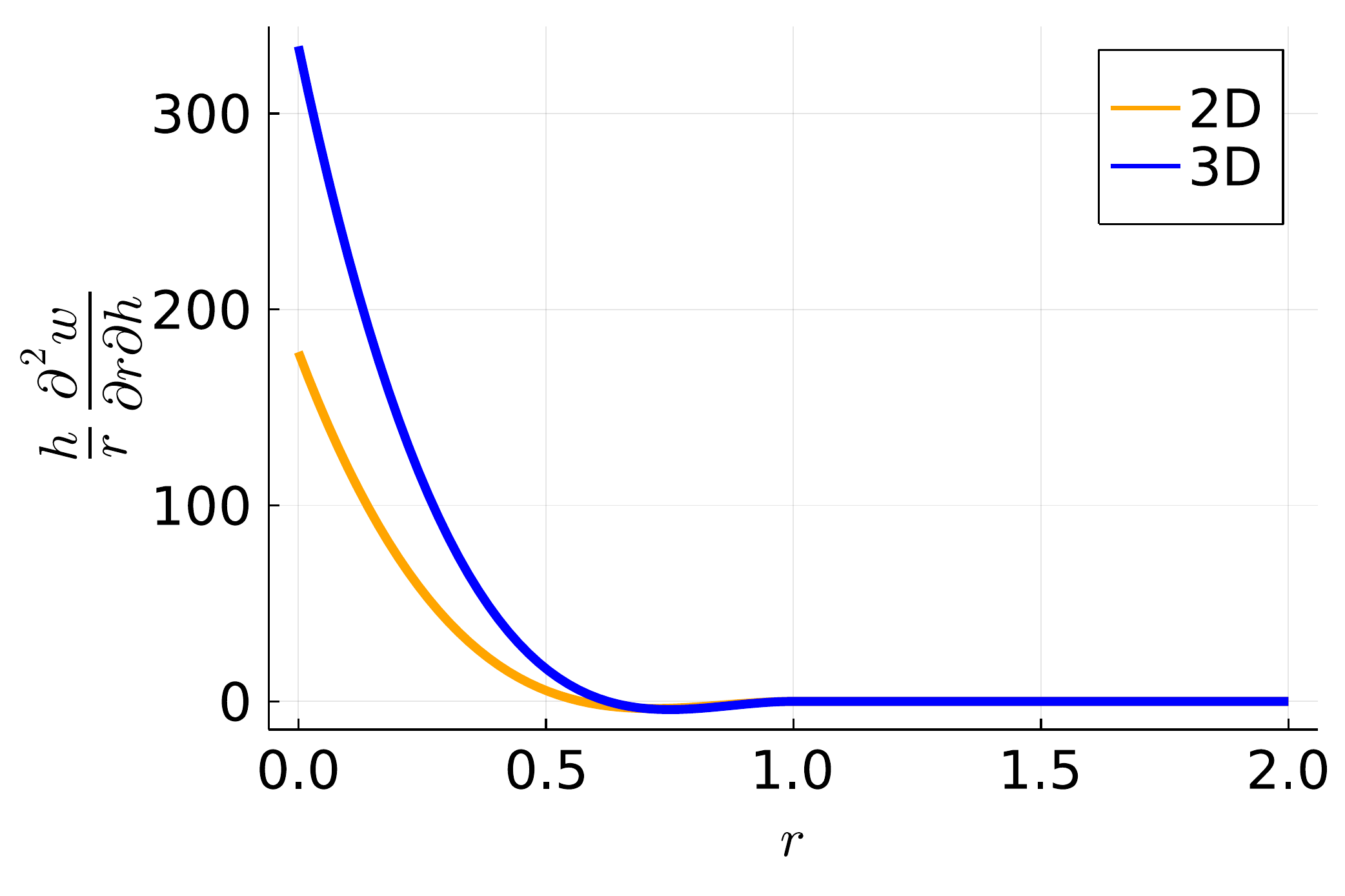}
	\caption{The plot of $h\pdv{w}{h}$ (left) and $\frac{h}{r} \frac{\partial^2 w}{\partial r \partial h}$ (right) in two and three dimensions for $h = 1$.}
	\label{fig:tp}
\end{figure}

From the graphs in Figure \ref{fig:tp}, we can intuitively understand the behavior of artificial force in \eqref{eq:bom_with_tp} as follows: for evenly distributed particles (such as in a grid), $\lambda_a$ will be close to zero, since this is the average value of $h\pdv{w}{h}$ inside the ball of radius $2h$ (or disc in 2D). However, $\lambda_a$ will be negative when the particle $a$ is found in a cluster or chain of particles surrounded by a void. This activates the artificial force, whose magnitude is proportional to $\frac{h}{r} \frac{\partial^2 w}{\partial r \partial h}$. This makes it strongly repulsive for nearby particles and slightly attractive for relatively large separations within the particle's sphere of influence. Thus, we get a modification of the equations similar to Monaghan's anti-clump term but with the additional benefit that the energy is conserved, albeit in a modified form, and, as we will see in the next section, the contribution of $\calP_h$ to the total energy is usually small.

\subsection{Time integrator}
So far, we have only been concerned with the spatial semi-discretization (discrete space, 
continuous time), and a time integrator is required to solve the SHTC-SPH ordinary differential 
equations (ODE). ODE system \eqref{eq:odes} conserve energy, and we would like to find a time 
integrator 
that 
preserves this property. First, let us write the system in a more succinct form:
\begin{equation}
	\begin{split}
		\dot{\xx}_a &= \vv_a, \\
		\dot{\vv}_a &= \frac{\ff_a}{m_a}, \\
		\dot{\bbA}_a &= -\bbA_a \bbL_a, \\ 
	\end{split}
	\label{eq:odes-brief}
\end{equation}
where the force
\begin{equation*}
	\begin{split}
		\ff_a = &\sum_{b} m_a m_b \bigg(\bbT_a(\xx, \rho, \bbA) + \bbT_b(\xx, \rho, \bbA)\bigg) \nabla w_{ab} \\
		&- \sum_{b}  m_a m_b c_p^2 \frac{\lambda_a + \lambda_b}{\rho_0^2} \, \frac{h}{r_{ab}} \frac{\partial^2 w_{ab}}{\partial r_{ab} \partial h} \xx_{ab}.
	\end{split}
\end{equation*}
depends on $\xx$, $\bbA$, $\rho$ and $\lambda$. However, from equations \eqref{eq:rho}, \eqref{eq:lambda}, we see that $\rho$ and $\lambda$ are themselves merely functions of $\xx$, and thus we have $\ff_a = \ff_a(\xx, \bbA)$. For the discrete velocity gradient, we can write $\bbL_a = \bbL_a(\xx, \vv)$. 

Naturally, we would like to use a symplectic integrator, such as the Verlet scheme, which has excellent energy-conservation properties \cite{hairer}. Unfortunately, our system is not symplectic, due to the presence of variable $\bbA_a$. Instead, we suggest the following combination of Verlet (for $\xx$ and $\vv$) and the mid-point rule (for $\bbA$):
\begin{equation}
	\begin{split}
		\vv_a\left(t_{k+\frac{1}{2}}\right) &= \vv_a(t_k) + \frac{\delta t}{2m_a} \ff_a(t_k),\\
		\xx_a\left(t_{k+\frac{1}{2}}\right) &= \xx_a(t_k) + \frac{\delta t}{2} \vv_a\left(t_{k+\frac{1}{2}}\right),\\
		\bbA_a(t_{k+1})          &= \bbA_a(t_k) \left( \bbI - \frac{\delta t}{2} \bbL_a\left(t_{k+\frac{1}{2}}\right) \right) \left( \bbI + \frac{\delta t}{2} \bbL_a\left(t_{k+\frac{1}{2}}\right) \right)^{-1},\\
		\xx_a(t_{k+1}) &= \xx_a\left(t_{k+\frac{1}{2}}\right) + \frac{\delta t}{2} \vv_a\left(t_{k+\frac{1}{2}}\right),\\
		\vv_a(t_{k+1}) &= \vv_a\left(t_{k+\frac{1}{2}}\right) + \frac{\delta t}{2m_a} \ff_a(t_{k+1}),
	\end{split}
	\label{eq:time-step}
\end{equation}
where $t_k = k \, \delta t$ denotes the $k$-th time-step, and $\ff_a(t)$, $\bbL_a(t)$ is a shorthand notation for
\begin{equation*}
	\ff_a(t) = \ff_a(\xx(t), \bbA(t)), \qquad \bbL_a(t) = \bbL_a(\xx(t), \vv(t)).
\end{equation*}
From the practical standpoint, this scheme is explicit in the sense that there are no linear or non-linear systems to be solved, or matrices to be inverted, except those of size $d\times d$. The main motivation for using \eqref{eq:time-step} is to obtain discrete time-reversibility as in \cite{sph-reversible}. Indeed, inverting the sign of $\vv_a$ and $\bbL_a$ in \eqref{eq:time-step}, we get the exactly same set of equations with the swapped role of $t_{k}$ and $t_{k+1}$.

\subsection{Adding relaxation}
We now have a discrete system for elastic solid in terms of arrays $\xx, \vv, \bbA$, which 
constitutes the reversible part of SHTC framework. The last step is adding fluidity to our model by 
relaxing $\bbA$, and hence tangential stresses. Let us return to semi-discrete differential system 
\eqref{eq:odes-brief}, where we add relaxation as\footnote{The function $ \theta $ in 
\eqref{eqn.PDE.A} is taken 
as $ \theta = \frac{\tau c_s^2}{3} (\det \bbA)^{-5/3}  $, see \cite{DPRZ2016}. However, we omit the 
factor $ (\det \bbA)^{-5/3} $ in this paper for simplicity because it is important only in the case 
of compressible viscous fluids which we shall not consider here.}
\begin{subequations}\label{eq:odes-brief-relax}
	\begin{align}
		\dot{\xx}_a &= \vv_a, \\
		\dot{\vv}_a &= \frac{\ff_a}{m_a}, \\
		\dot{\bbA}_a &= -\bbA_a \bbL_a - \frac{3}{\tau c_s^2} \epsilon_{\bbA_a},
	\end{align}
\end{subequations}
where $\tau$ is the \textit{relaxation time} (noting that the potential equilibrium of $\bbA$ is 
$\bbA =
\mathbb{Q}$ with $ \mathbb{Q} $ being an orthogonal matrix). For an elastic solid $ \tau = \infty $ 
 and for a Newtonian fluid $\tau$ is  
a constant, while for non-Newtonian fluids and elastoplastic solids it should be taken as a 
function of $\bbA_a$ \cite{nonNewtonian2021,Hyper-Hypo2019}. With this new addition, the equation 
for $\bbA_a$ is often stiff, and therefore, implicit and exponential time integrators are 
recommended \cite{DPRZ2016,LGPR2022}. However, we already have a very small time step in 
our explicit SHTC-SPH integrator (as opposed to fully implicit finite element or finite volume 
approaches), so for 
performance reasons, we use a simpler splitting strategy, using the classical RK4 scheme 
\cite{hairer} for integrating the relaxation term in the PDE for $\bbA_a$:
\begin{subequations}	\label{eq:time-step-relax}
	\begin{align}
		\vv_a\left(t_{k+\frac{1}{2}}\right) &= \vv_a(t_k) + \frac{\delta t}{2m_a} \ff_a(t_k),\\
		\xx_a\left(t_{k+\frac{1}{2}}\right) &= \xx_a(t_k) + \frac{\delta t}{2} \vv_a\left(t_{k+\frac{1}{2}}\right),\\
		\bbA_a^*(t_{k})          &= \bbA_a(t_k) \left( \bbI - \frac{\delta t}{2} \bbL_a\left(t_{k+\frac{1}{2}}\right) \right) \left( \bbI + \frac{\delta t}{2} \bbL_a\left(t_{k+\frac{1}{2}}\right) \right)^{-1},\\
		\bbA_a(t_{k+1})          &= \bbA_a^*(t_{k+1})   + \frac{\delta t}{6} \bigg( \bbK_{1,a}(t_{k}) + 2\bbK_{2,a}(t_{k}) +2\bbK_{3,a}(t_{k}) + \bbK_{4,a}(t_{k}) \bigg),\\
		\xx_a(t_{k+1}) &= \xx_a\left(t_{k+\frac{1}{2}}\right) + \frac{\delta t}{2} \vv_a\left(t_{k+\frac{1}{2}}\right),\\
		\vv_a(t_{k+1}) &= \vv_a\left(t_{k+\frac{1}{2}}\right) + \frac{\delta t}{2m_a} \ff_a(t_{k+1}),
	\end{align}
\end{subequations}
where
\begin{equation*}
	\bbK_{i,a}(t_k) = -\left.  \frac{3}{\tau c_s^2} \epsilon_{\bbA_a} \right|_{\bbA = \bbA^*(t_k) + b_i \, \delta t \, \bbK_{i-1, a}(t_k)}, \qquad i = 1,2,3,4
\end{equation*}
and $(b_1, b_2, b_3, b_4) = (1, \frac{1}{2}, \frac{1}{2}, 1)$. Therefore, we get relatively cheap 
time steps and for $\tau = \infty$ (no relaxation) the scheme reduces to the reversible one 
\eqref{eq:time-step}. 

Since positions are updated twice per step, we also require two neighbor list calculations in each iteration of \eqref{eq:time-step}.

\section{Numerical results}
As we have introduced the new SHTC-SPH numerical scheme, the following section contains the 
numerical results for both fluids and solids to demonstrate the robustness of the proposed SHTC-SPH 
approach. For a list of all the material and SPH parameters used in 
different test cases, we refer to Table \ref{table:parameters}.

\begin{table}[hbt!]
	\begin{tabular}{|l|l|l|l|l|l|l|l|}
		\hline
		& model & $\dd{r}$       & $\dd{t}$       & $c_0$     & $c_s$     & $c_p$     & $h$        
		\\
		\hline
		Beryllium plate           & DPR   & 2.50E-04 & 9.05E-10 & 9.05E+03 & 9.05E+03 & 9.05E+04 & 
		3.75E-04 \\
		\hline
		Twisting column           & NH    & 4.96E-02 & 3.44E-06 & 7.15E+02 & 7.19E+01 & 7.15E+03 & 
		7.44E-02 \\
		\hline
		Taylor-Couette flow       & DPR   & 3.33E-02 & 3.31E-05 & 2.00E+01 & 4.00E+01 & 2.00E-01 & 
		5.00E-02 \\
		\hline
		Lid-driven cavity, Re 100 & DPR   & 5.00E-03 & 1.50E-05 & 2.00E+01 & 2.00E+01 & 0.00E+00 & 
		7.50E-03 \\
		\hline
		Lid-driven cavity, Re 400 & DPR   & 7.14E-03 & 1.50E-05 & 2.00E+01 & 2.00E+01 & 0.00E+00 & 
		1.07E-02 \\
		\hline
	\end{tabular}
	\caption{Summary of simulation parameters used in this paper.}
	\label{table:parameters}
\end{table}

\subsection{Beryllium plate}
\begin{figure}[!htb]
	\centering
	\includegraphics[draft=false,width=0.55\textwidth]{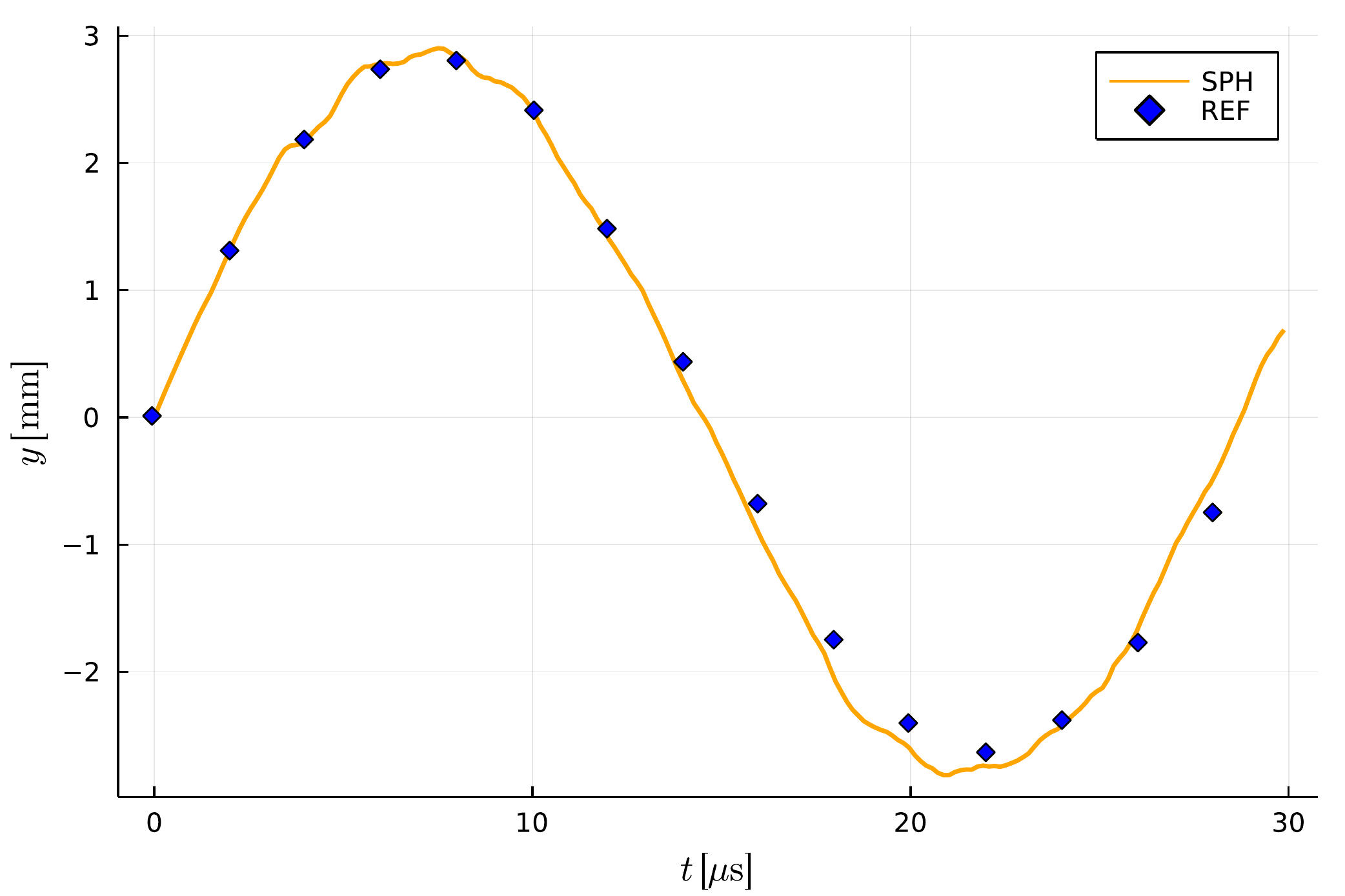}
	\caption{The $y$ coordinate of center point in beryllium plate benchmark plotted against time. Orange line depicts the result of our scheme. Blue squares mark the referential solution.}
	\label{fig:plate-oscillations}
\end{figure}
\begin{figure}[!htb]
	\centering
	\includegraphics[draft=false,width=0.55\textwidth]{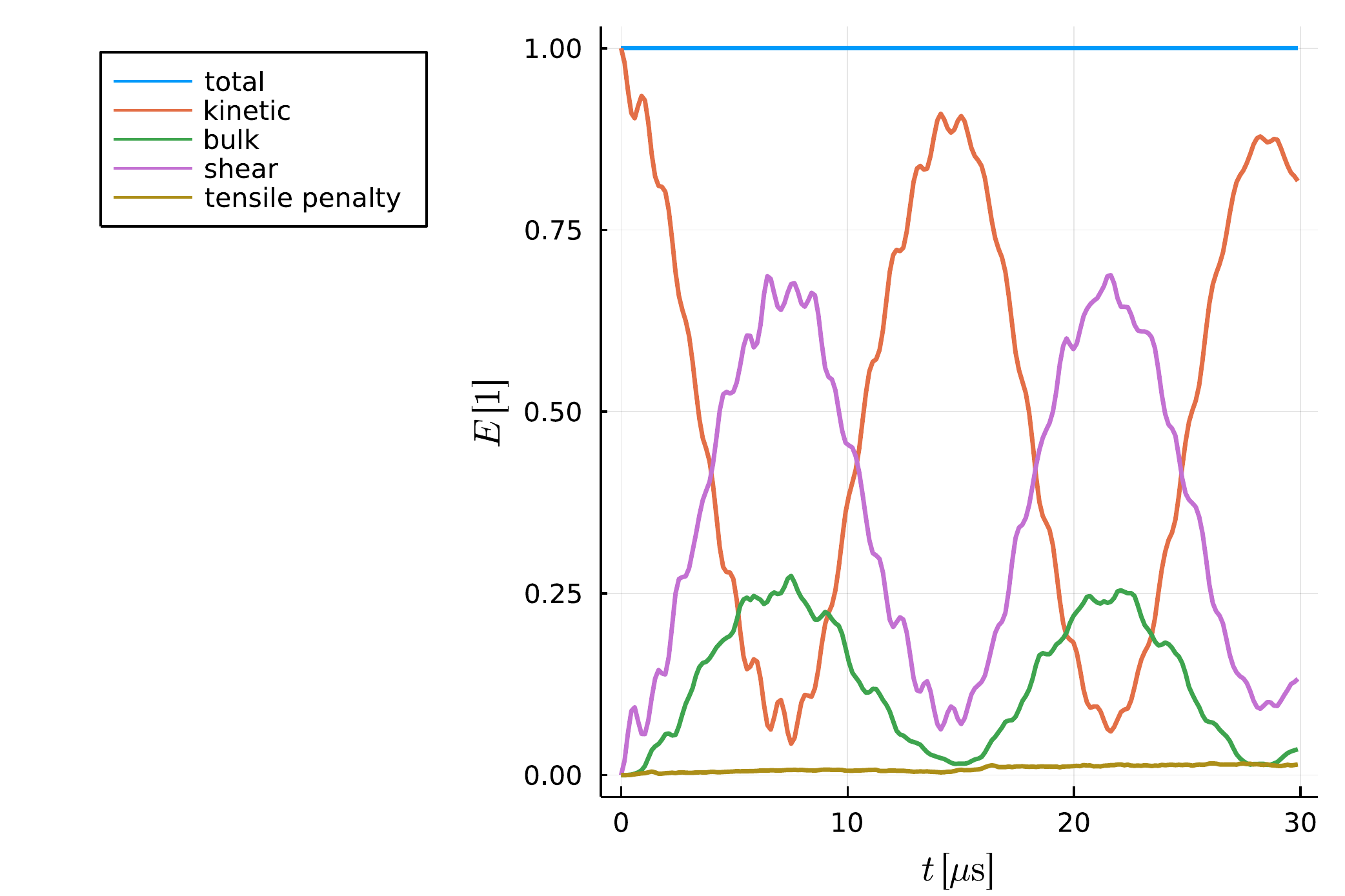}
	\caption{Different contributions to energy in beryllium plate benchmark plotted against time, normalized. Interestingly, the tensile penalty contribution is kept relatively small despite its profound importance in the evolution.}
	\label{fig:plate-energy}
\end{figure}
\begin{figure}[!htb]
	\begin{subfigure}{0.49\textwidth}
		\includegraphics[draft=false,width=\textwidth, clip]{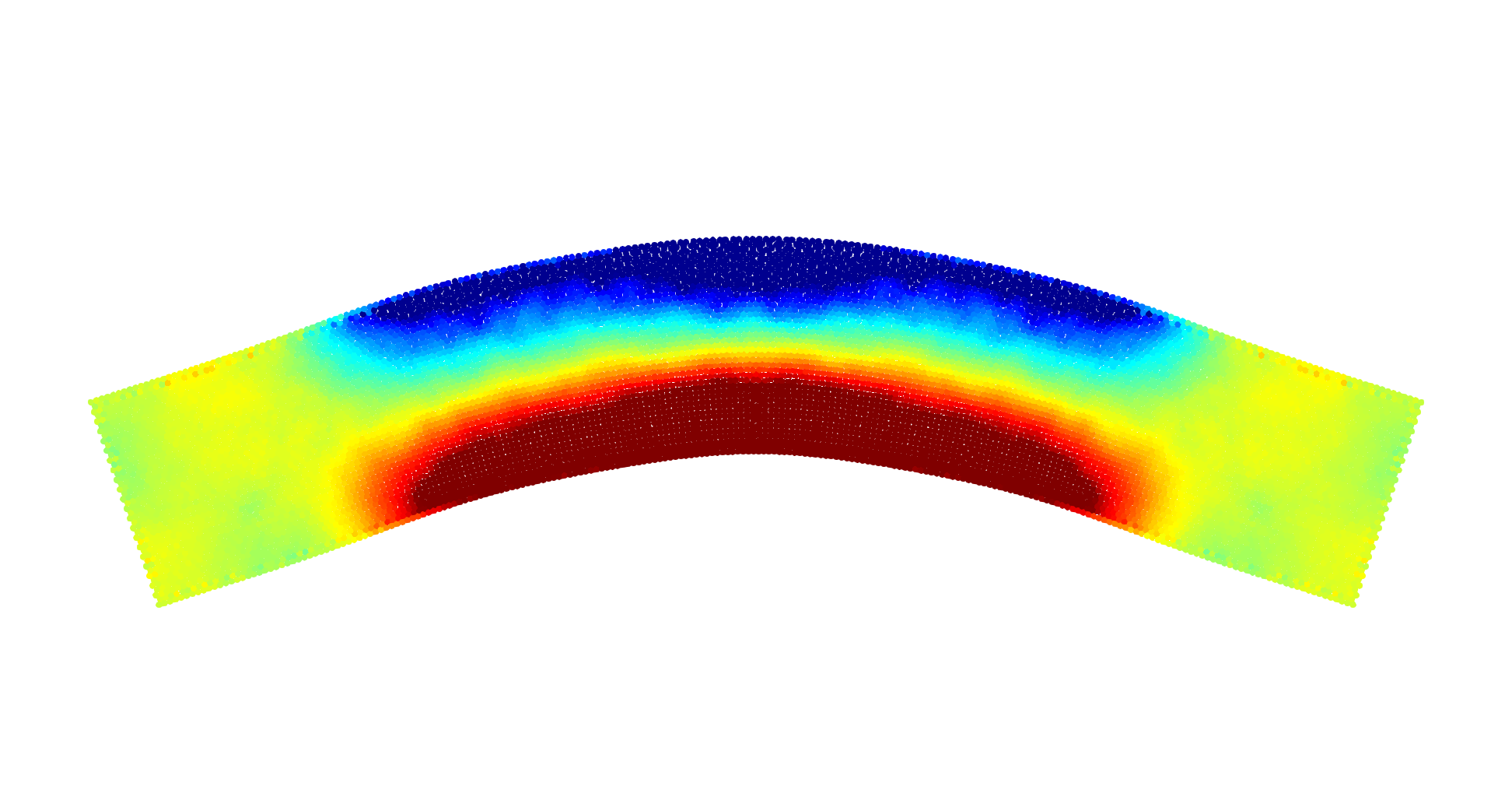}
		\caption{$t = 0.075\si{\mu s}$, $c_p/c_0 = 10$}
	\end{subfigure}
	\begin{subfigure}{0.49\textwidth}
		\includegraphics[draft=false,width=\textwidth, clip]{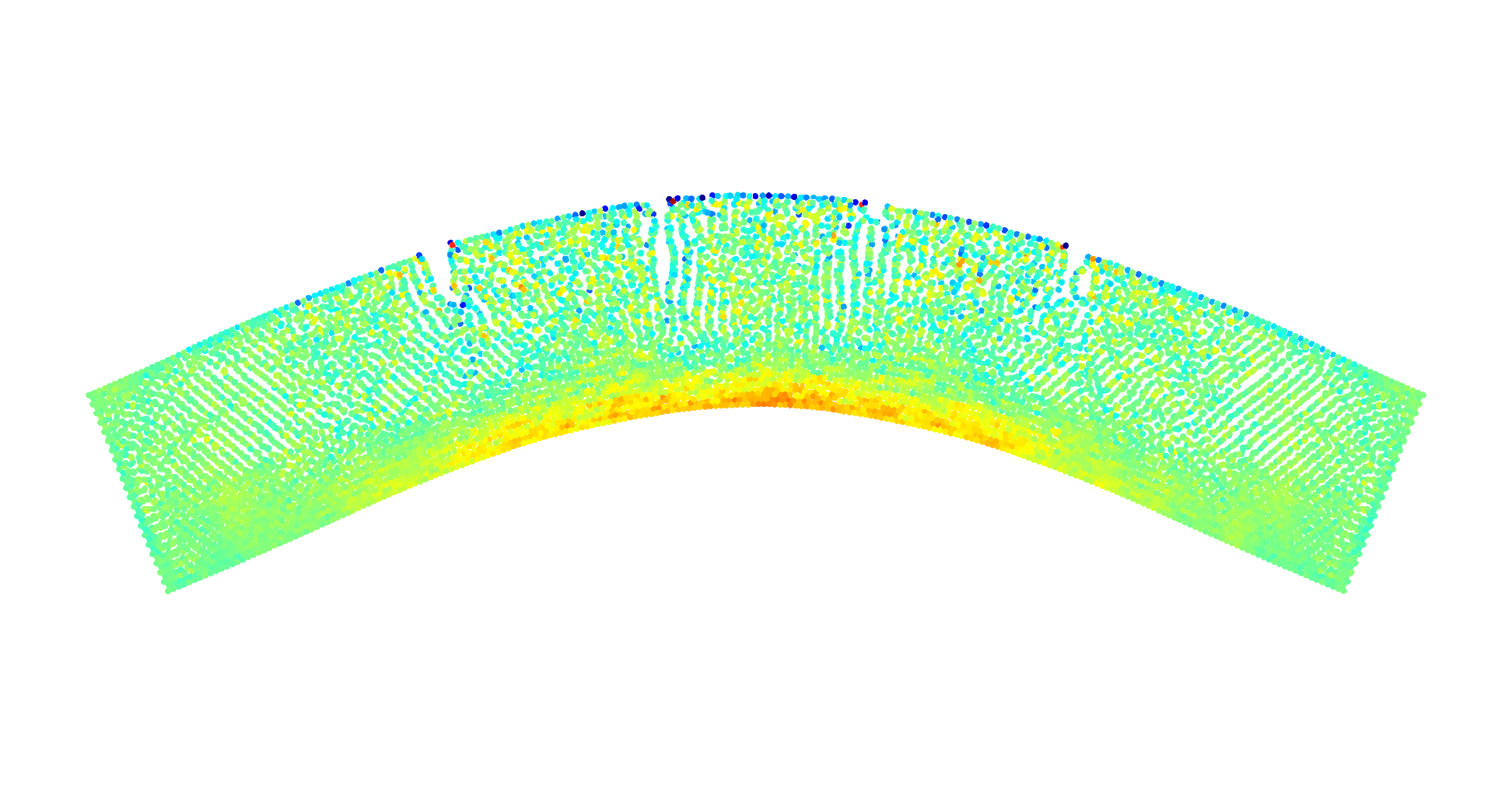}
		\caption{$t = 0.075\si{\mu s}$, $c_p/c_0 = 0$}
	\end{subfigure}
	\begin{subfigure}{0.49\textwidth}
		\includegraphics[draft=false,width=\textwidth, clip]{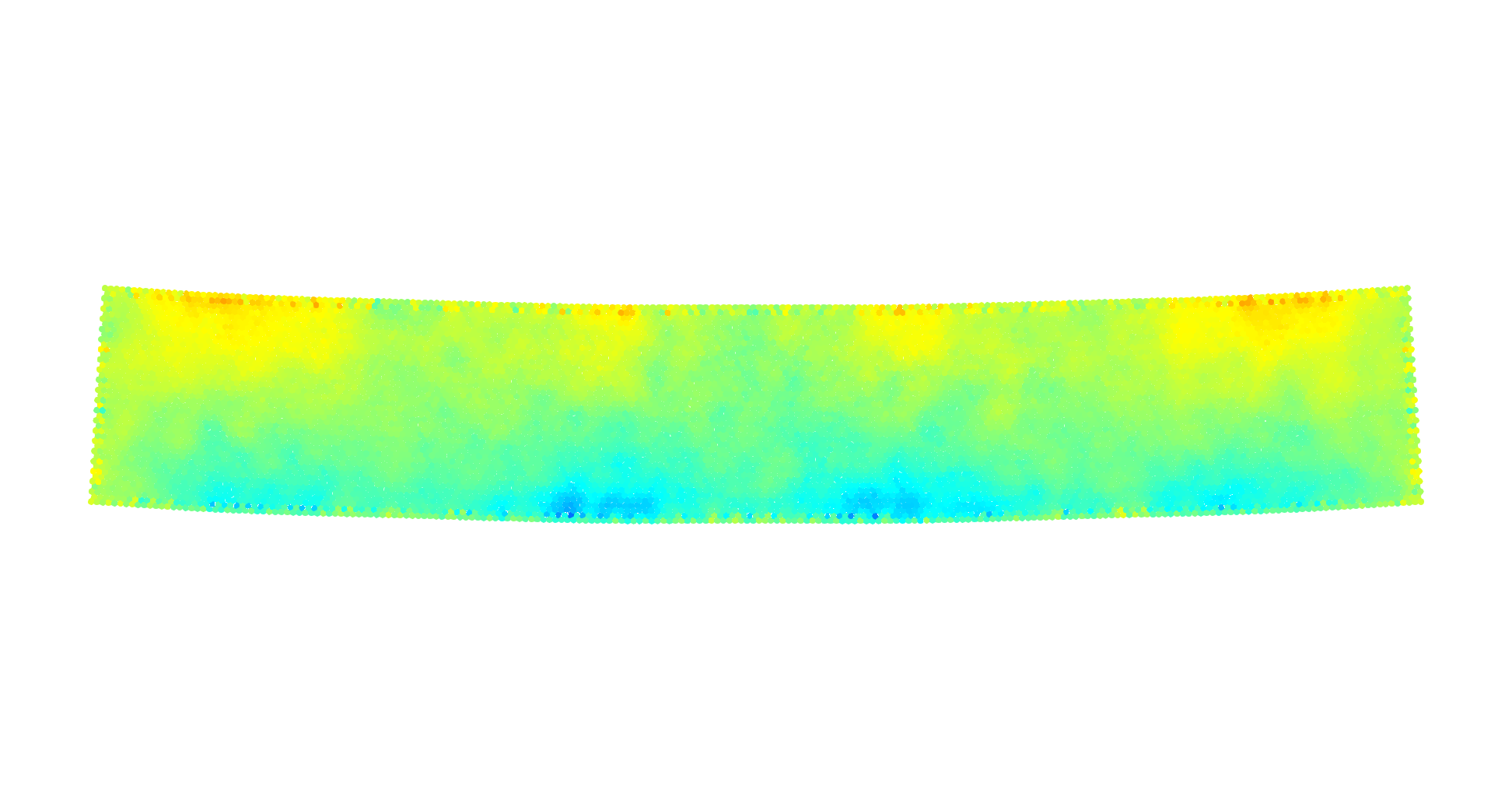}
		\caption{$t = 0.150\si{\mu s}$, $c_p/c_0 = 10$}
	\end{subfigure}
	\begin{subfigure}{0.49\textwidth}
		\includegraphics[draft=false,width=\textwidth, clip]{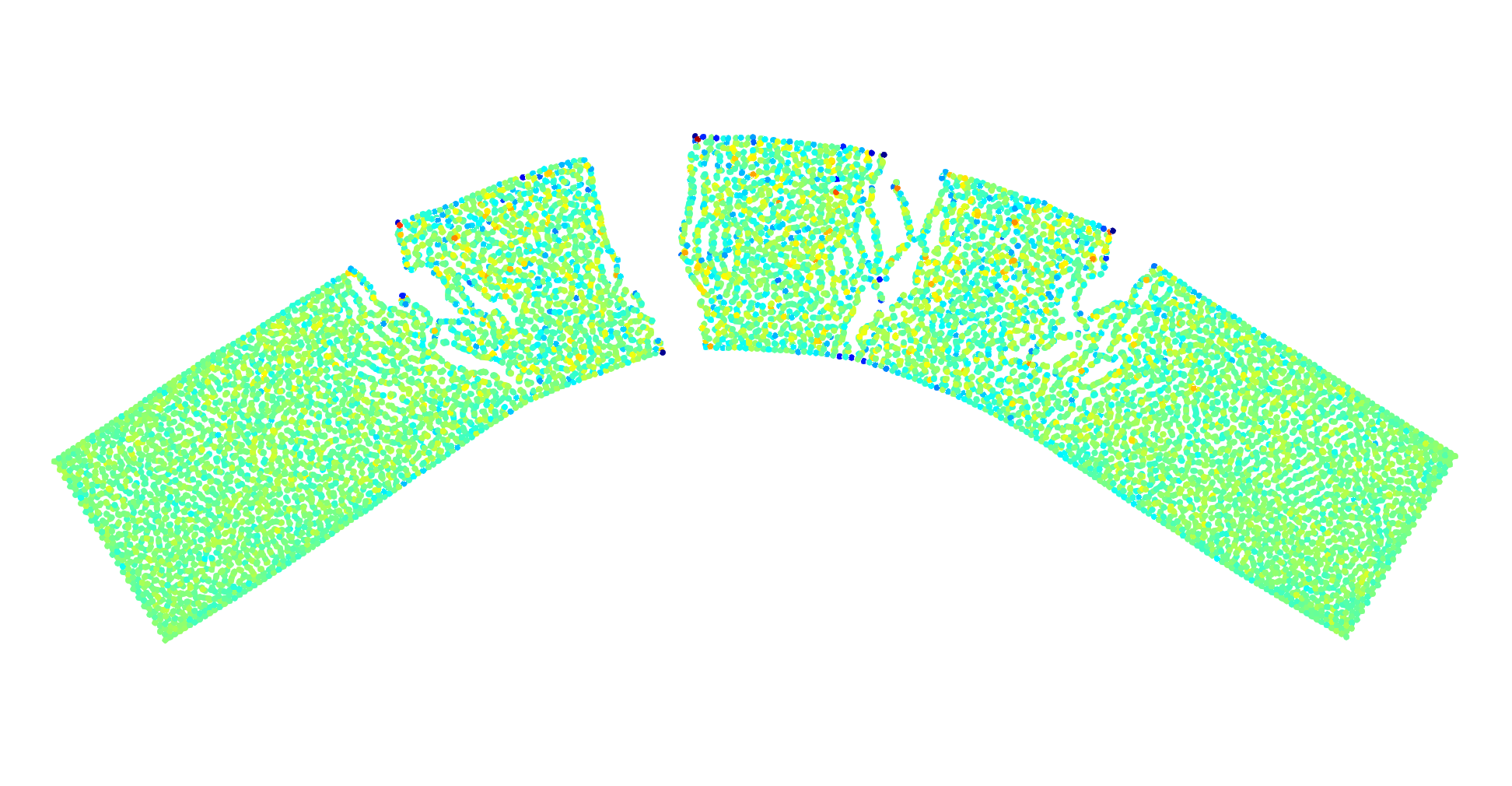}
		\caption{$t = 0.150\si{\mu s}$, $c_p/c_0 = 0$}
	\end{subfigure}
	\begin{subfigure}{0.49\textwidth}
		\includegraphics[draft=false,width=\textwidth, clip]{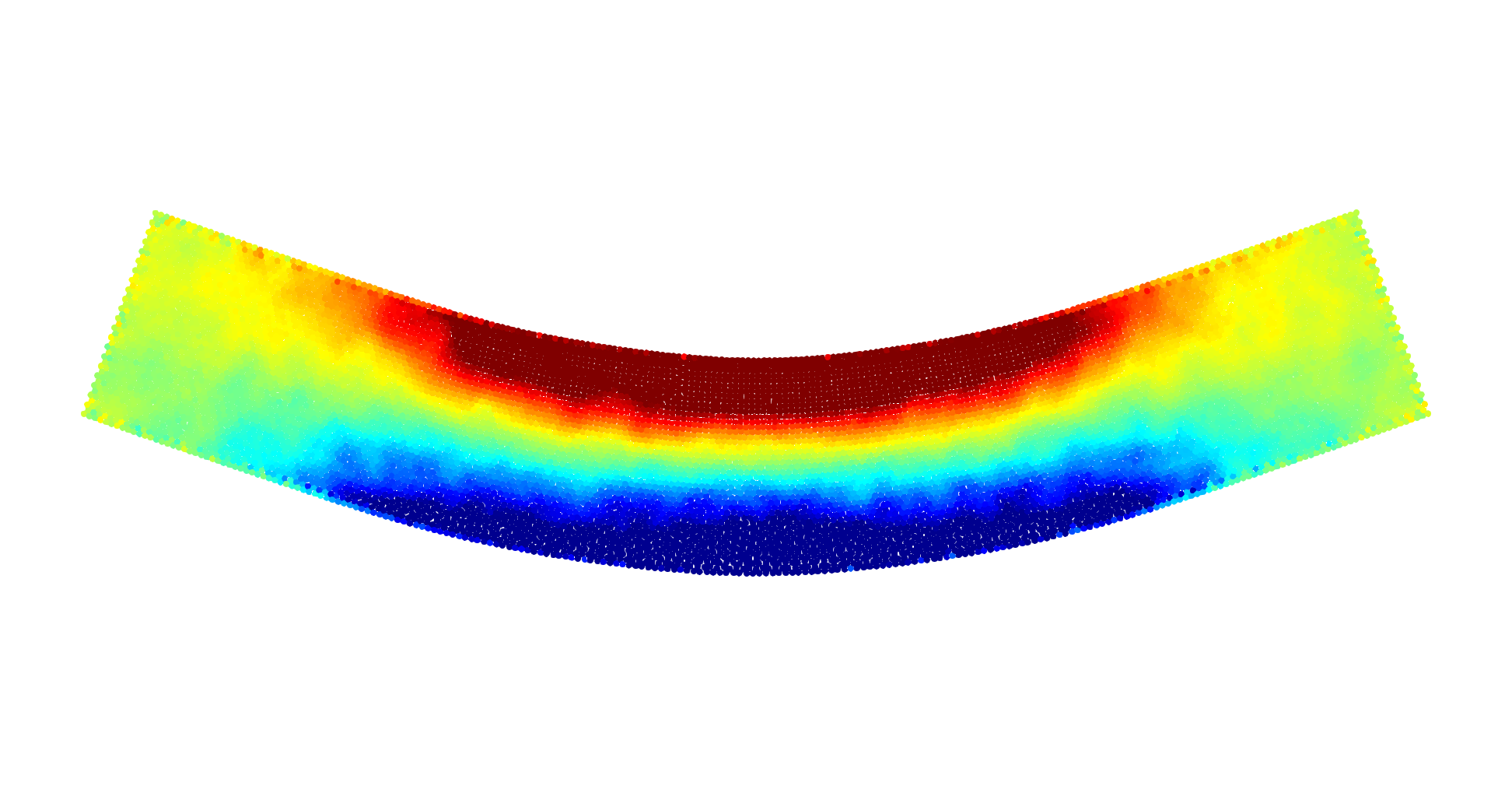}
		\caption{$t = 0.225\si{\mu s}$, $c_p/c_0 = 10$}
	\end{subfigure}
	\begin{subfigure}{0.49\textwidth}
		\includegraphics[draft=false,width=\textwidth, clip]{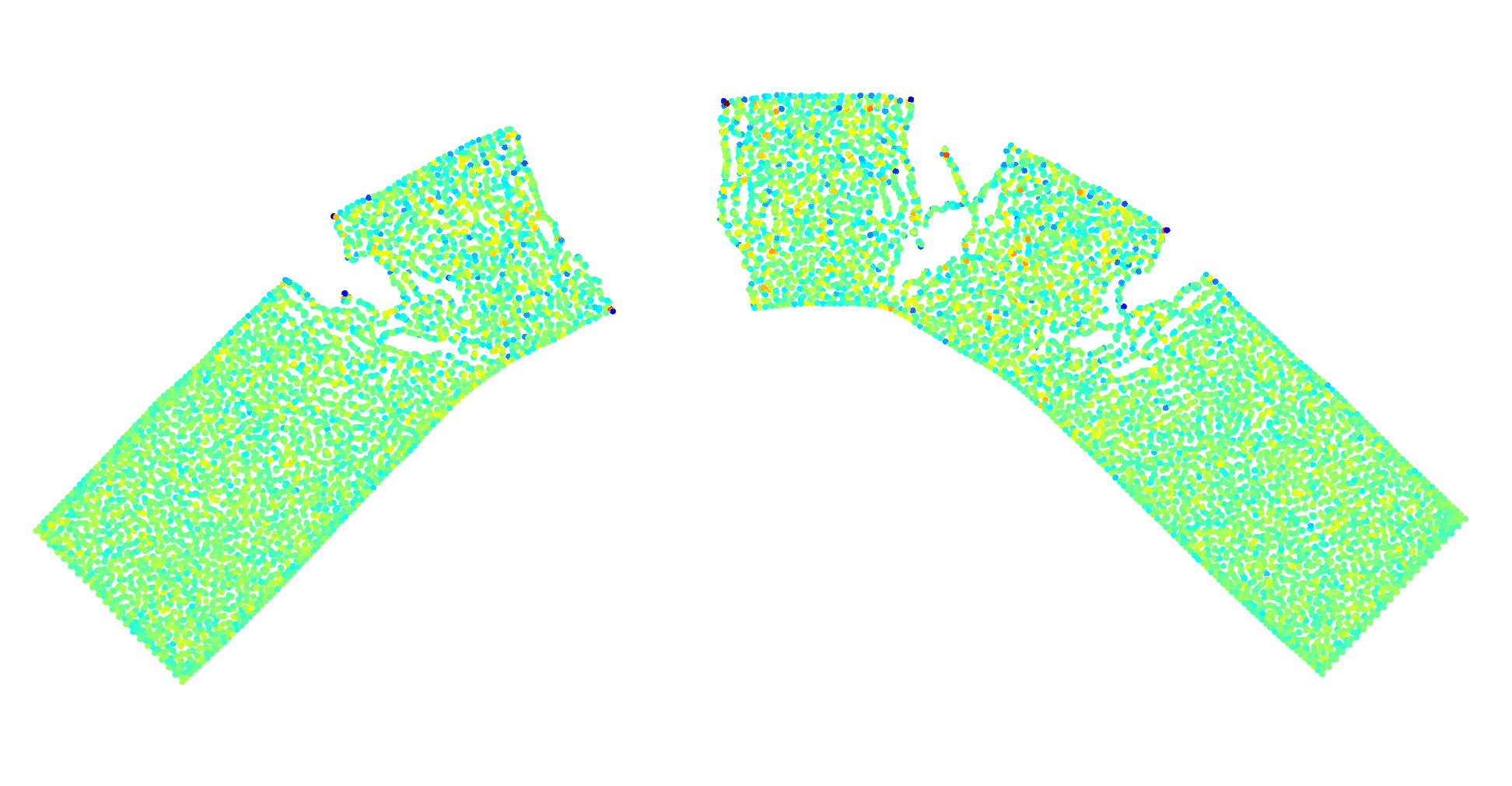}
		\caption{$t = 0.225\si{\mu s}$, $c_p/c_0 = 0$}
	\end{subfigure}
	\centering
	\includegraphics[draft=false,width=0.5\textwidth]{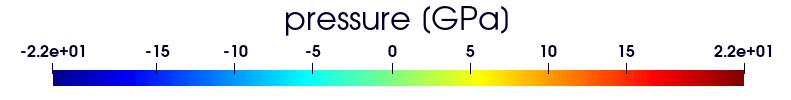}
	\caption{Images of beryllium plate simulation and color-plot for pressure with tensile 
	instability treatment (left) and without (right). For $c_p = 0$, particles have an extra degree 
	of freedom which allows them to minimize negative pressure by forming clumps. This results in 
	clearly non-physical behavior.}
	\label{fig:plate-pics}
\end{figure}
This benchmark examines the two-dimensional oscillation of an elastic solid that is bending due to a velocity field  prescribed at $t=0$. The body in question is a translationally symmetric plate, whose cross section is a rectangle 
$$ \Omega = \left( -\frac{L}{2}, \frac{L}{2} \right) \times \left(-\frac{W}{2}, \frac{W}{2} \right)$$
with values $L = 0.06 \, \si{m}$ and $W = 0.01 \, \si{m}$. The initial velocity field in this cross section is
$$ \vv = A \omega \begin{pmatrix}
	0 \\
	a_1 \left(\sinh s + \sin s\right) - a_2 \left(\cosh s + \cos s \right),
\end{pmatrix}$$
where $s = \alpha \left(x + \frac{L}{2}\right)$ and $A, \omega, a_1, a_2, \alpha$ are constants 
with values (parameters retrieved from \cite{LGPR2022}):
\begin{equation*}
	\begin{split}
		A &= 4.3369 \cdot 10^{-5} \; \si{m},\\
		\omega &= 2.3597 \cdot 10^{5}  \; \si{s^{-1}},\\
		\alpha &= 78.834,\\
		a_1 &= 56.6368,\\
		a_2 &= 57.6455.
	\end{split}
\end{equation*}
We use the constitutive relation \eqref{eq:eps0}-\eqref{eq:DPR} for with 
$$c_0 = c_s = 9046.59 \, \si{m.s^{-1}}.$$
Complete elasticity is assumed and therefore $\tau = \infty$. We choose the spatial step $\delta r = W/40$ which gives us approximately $10 000$ particles. The time step is selected according to
\begin{equation}
	\delta t = \frac{0.05\delta r}{\sqrt{c_0^2 + \frac{4}{3}c_s^2}}
	\label{eq:time-step-size}
\end{equation}
and the simulation ends at time $t = 3\cdot 10^{-5} \, \si{s}$, roughly corresponding to one period 
of the oscillating motion. 

Three things can be tested in this benchmark. Firstly, we plot the $y$ coordinate of the central 
point, which we then compare to data from the finite volume simulation \cite{LGPR2022}. Figure 
\ref{fig:plate-oscillations} shows that we get a reasonable agreement. Second, since there are no 
dissipative or external forces involved, it presents an ideal test for verifying the conservation 
of energy, which we achieve to a reasonable degree, see Figure \ref{fig:plate-energy}. Last but 
not least, due to the presence of strongly negative pressures, this simulation poses a challenge 
with respect to tensile instability, demonstrating the usefulness of the penalty term 
\eqref{eq:penalty}. In fact, without this addition, the plate would tear completely as can be seen 
in Figure \ref{fig:plate-pics}. 

\subsection{Twisting column}
\begin{figure}[!hbt]
	\begin{subfigure}{0.24\textwidth}
		\includegraphics[draft=false,width=\textwidth, clip]{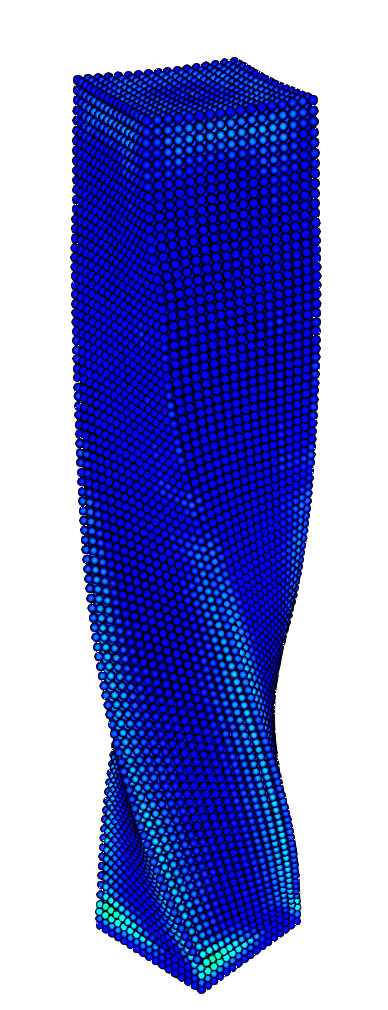}
		\caption{$t = 0.02\si{s}$}
	\end{subfigure}
	\begin{subfigure}{0.24\textwidth}
		\includegraphics[draft=false,width=\textwidth, clip]{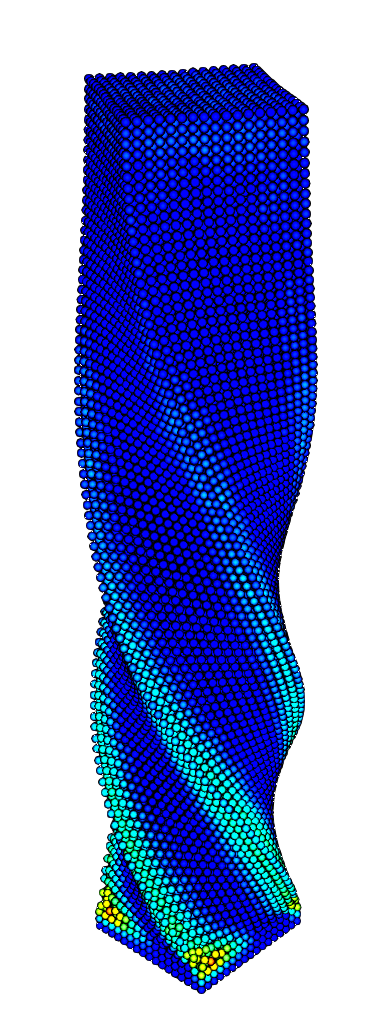}
		\caption{$t = 0.04\si{s}$}
	\end{subfigure}
	\begin{subfigure}{0.24\textwidth}
		\includegraphics[draft=false,width=\textwidth, clip]{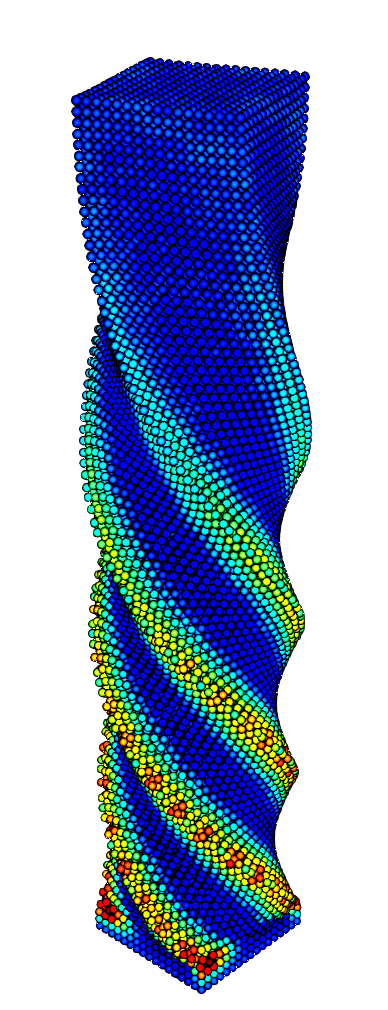}
		\caption{$t = 0.8\si{s}$}
	\end{subfigure}
	\begin{subfigure}{0.24\textwidth}
		\includegraphics[draft=false,width=\textwidth, clip]{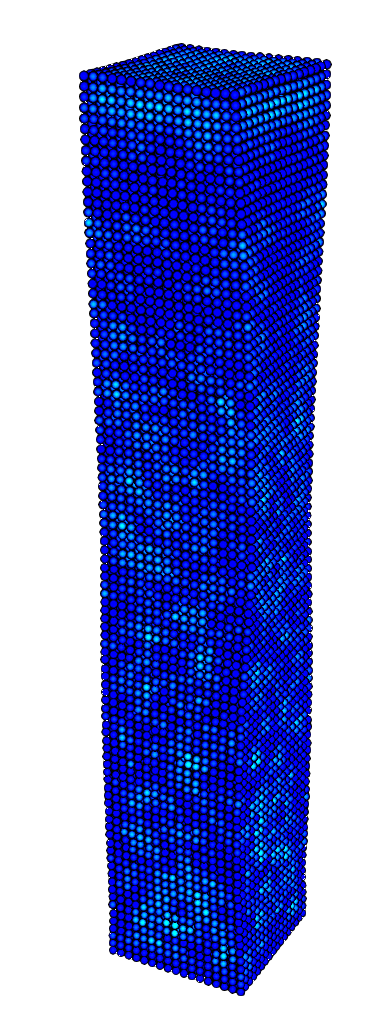}
		\caption{$t = 0.18\si{s}$}
	\end{subfigure}
	\centering
	\includegraphics[draft=false,width=0.5\textwidth]{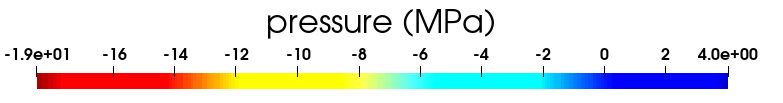}
	\caption{Result of the twisting column simulation with a color-plot for pressure. We use this as a qualitative test whether our code can deal with large elastic deformations in three dimensions. The last frame shows the column after untwisting, showing some noise in the pressure field (presumably, this can be remedied by adding artificial dissipation). Shape of the column is not exactly recovered but note that the elastic energy is not perfectly zero at this point either (see Figure \ref{fig:twist-energy}).}
	\label{fig:twist-pics}
\end{figure}
\begin{figure}[!htb]
	\centering
	\includegraphics[draft=false,width=0.5\textwidth]{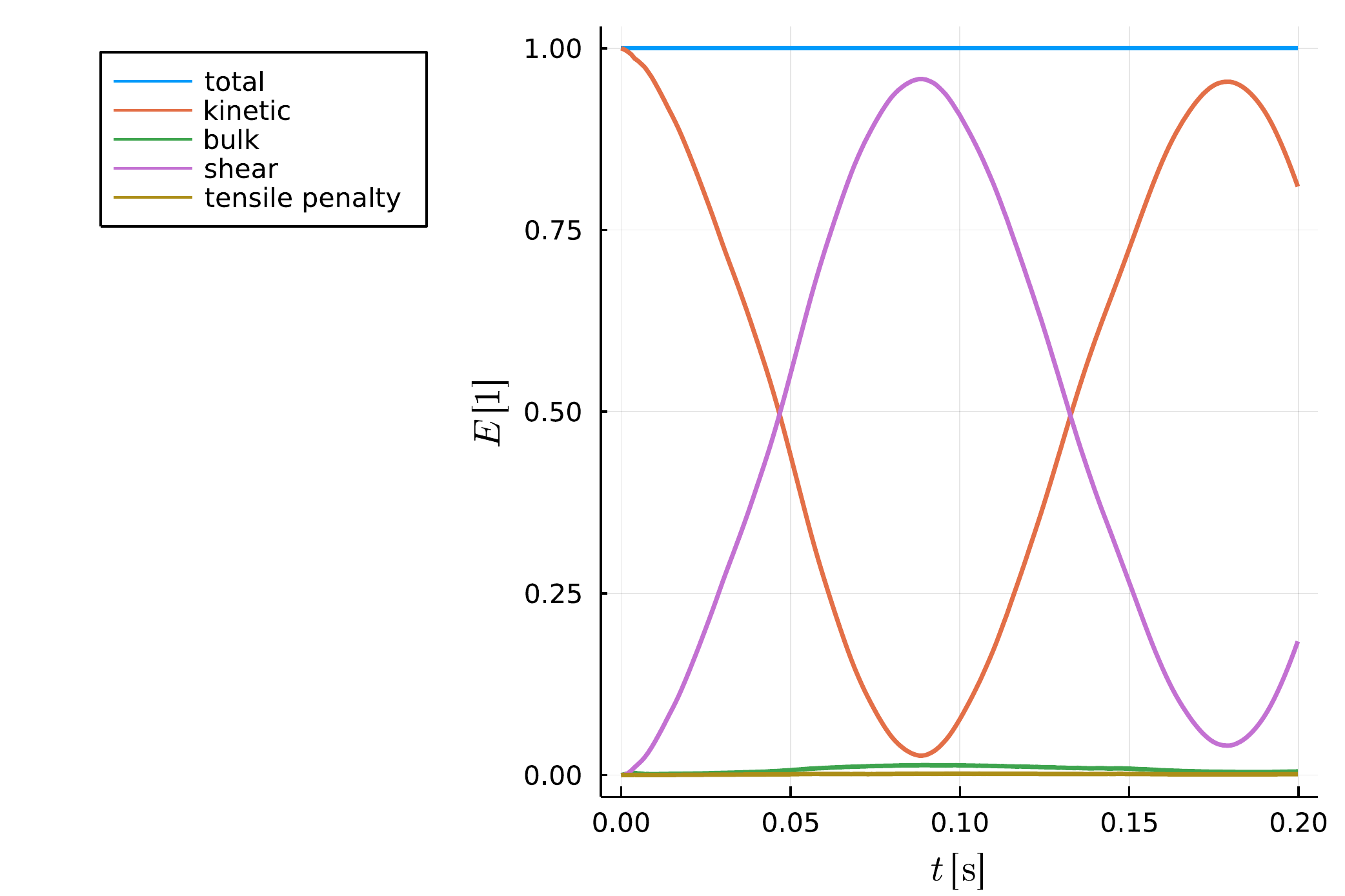}
	\caption{Energy conservation in twisting column benchmark.}
	\label{fig:twist-energy}
\end{figure}
\begin{figure}[!htb]
	\centering
	\includegraphics[draft=false,width=0.55\textwidth]{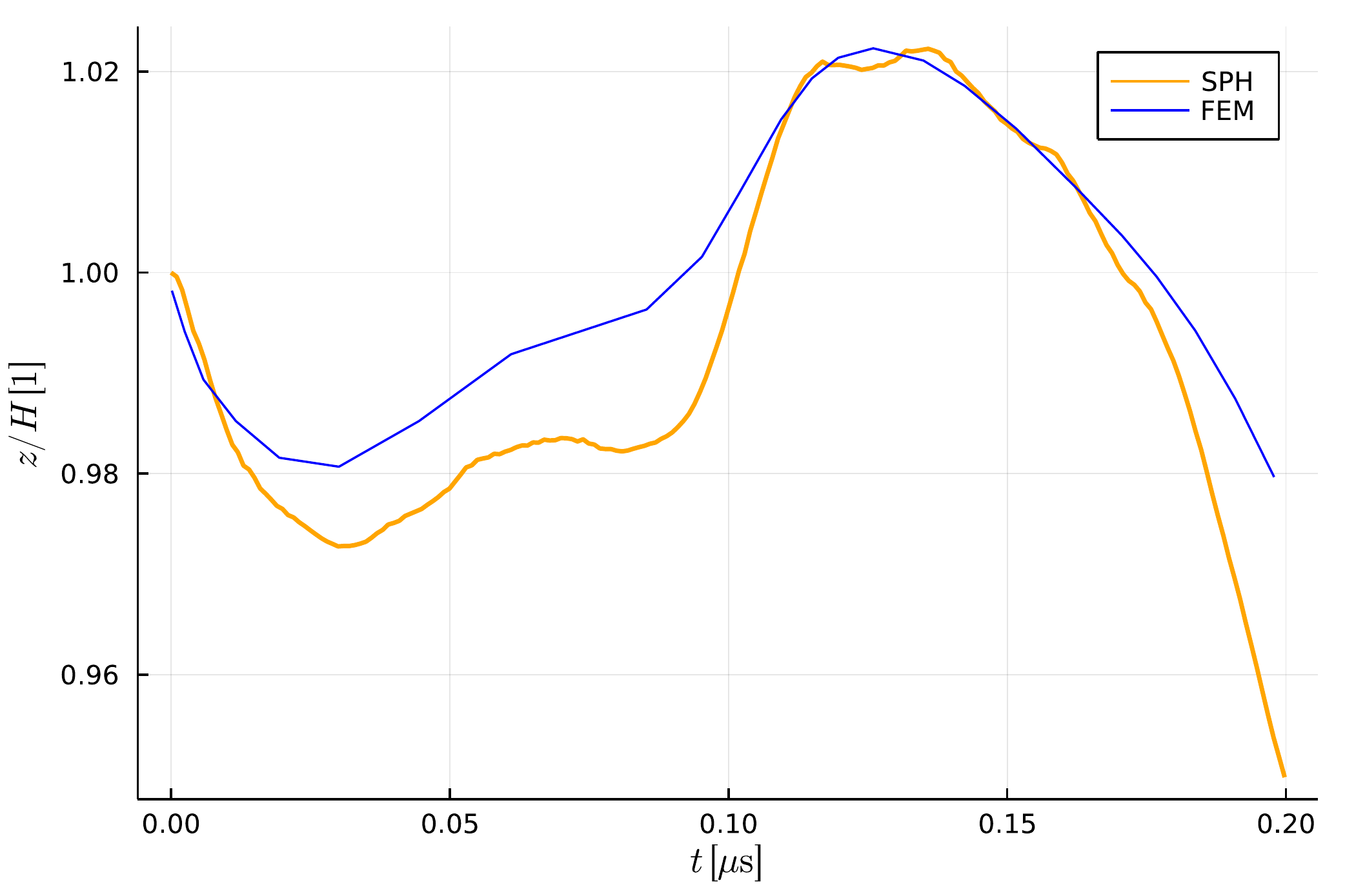}
	\caption{Twisting column: the normalized $z$-coordinate of material point initially at 
	$(0,0,H)$ plotted against time. Although the column is shrinking at first, it ``bounces off'' 
	at some point and starts to elongate. This effect was also observed in \cite{LGPR2022} and 
	\cite{haider2018upwind}. The blue line shows comparison to a finite element simulation (created 
	using the Fenics software \cite{fenics}).}
	\label{fig:twist-height}
\end{figure}
The next benchmark is borrowed from \cite{haider2018upwind}. The initial setup is a cuboid
\begin{equation}
	\Omega = \left(-\frac{W}{2}, \frac{W}{2} \right) \times \left(-\frac{W}{2}, \frac{W}{2}\right) \times \left(0, H\right)
\end{equation}
where $W = 1 \si{m}$ and $H = 6 \si{m}$ which is subjected to a prescribed velocity field
$$ \vv = \omega \sin \left(\frac{\pi z}{2 H}\right)\begin{pmatrix*}[r]
	y\\
	-x\\
	0
\end{pmatrix*}, $$
where $\omega = 105 \, \si{s ^{-1}}$. The base of the column is kept in place by a Dirichlet boundary condition for velocity. Here, we use the fully elastic ($\tau = \infty$) Neo-Hookean model \eqref{eq:NH} with $\rho_0 = 1100 \si{kg \per m^3}$, $Y = 17 \si{M Pa}$ (Young modulus), $\nu = 0.495$ (Poisson ratio). These values are related to the bulk shear sound speed by relations:
$$ \rho_0 c_s^2 = \frac{Y}{1 + \nu}, \qquad \rho_0 c_0^2 = \frac{\nu Y}{\left(1 + \nu\right) \left(1 - 2\nu\right)}.$$

In this benchmark, inertia should twist the column, building up tensile forces that eventually prevail and reverse the rotation. The shape of the column should recover without loss of energy. Additionally, there is associated non-linear effect, which causes shrinkage of the column. All these phenomena are observed in our simulation, as can be seen in Figures \ref{fig:twist-pics}, \ref{fig:twist-height}. We still have reasonably good energy conservation (Figure \ref{fig:twist-energy}), but the Dirichlet boundary condition at the base is slightly dissipative (since it is implemented by resetting the velocity to zero at every time step). Unfortunately, variables like pressure become very noisy in the simulation after a short time, but we did not find a remedy which would not involve artificial dissipation.

\subsection{Laminar Taylor-Couette flow}
\begin{figure}
	\begin{subfigure}{0.49\textwidth}
		\includegraphics[draft=false,width = \textwidth]{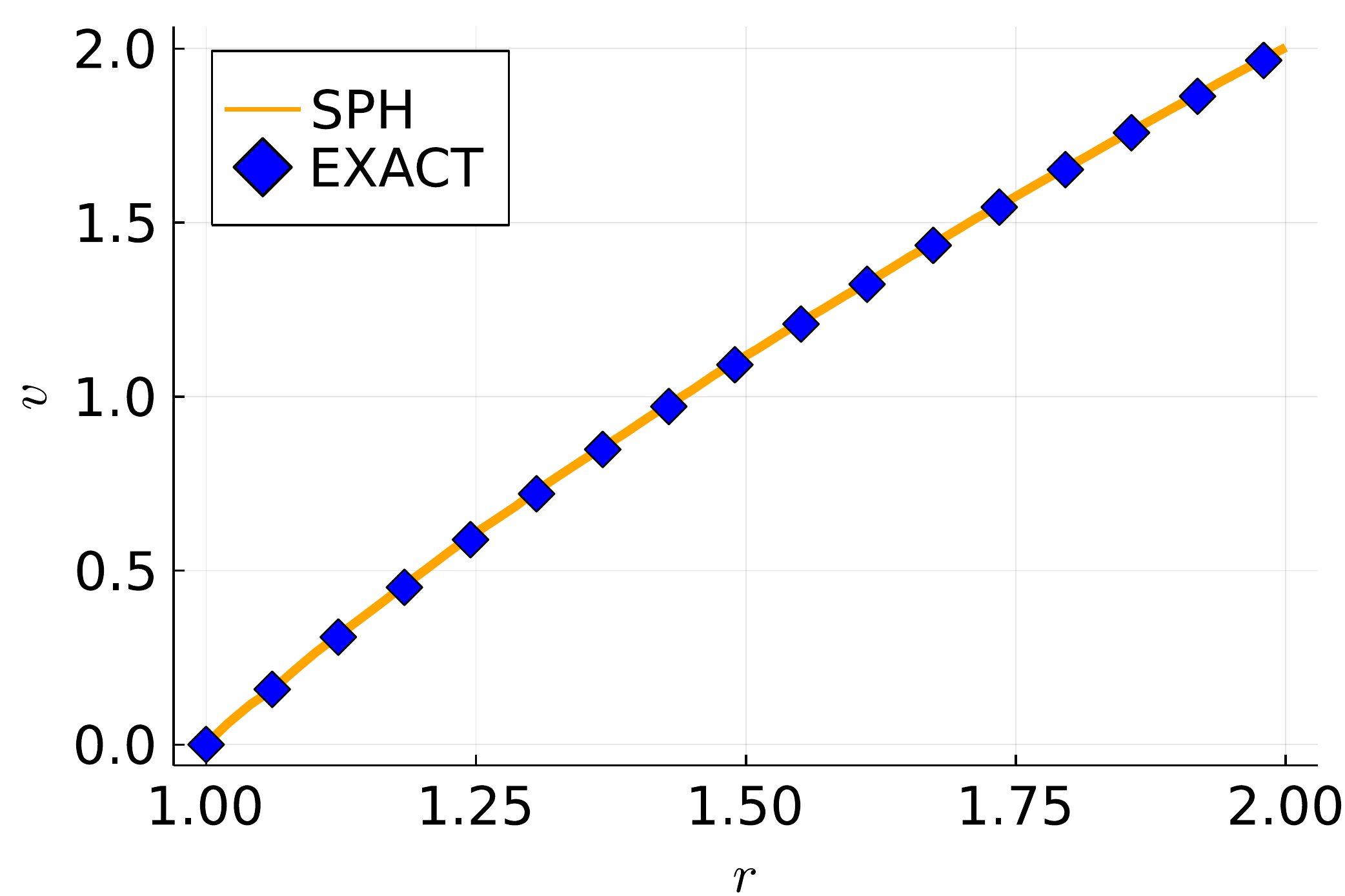}
		\caption{$v_y$}
	\end{subfigure}
	\begin{subfigure}{0.49\textwidth}
		\includegraphics[draft=false,width = \textwidth]{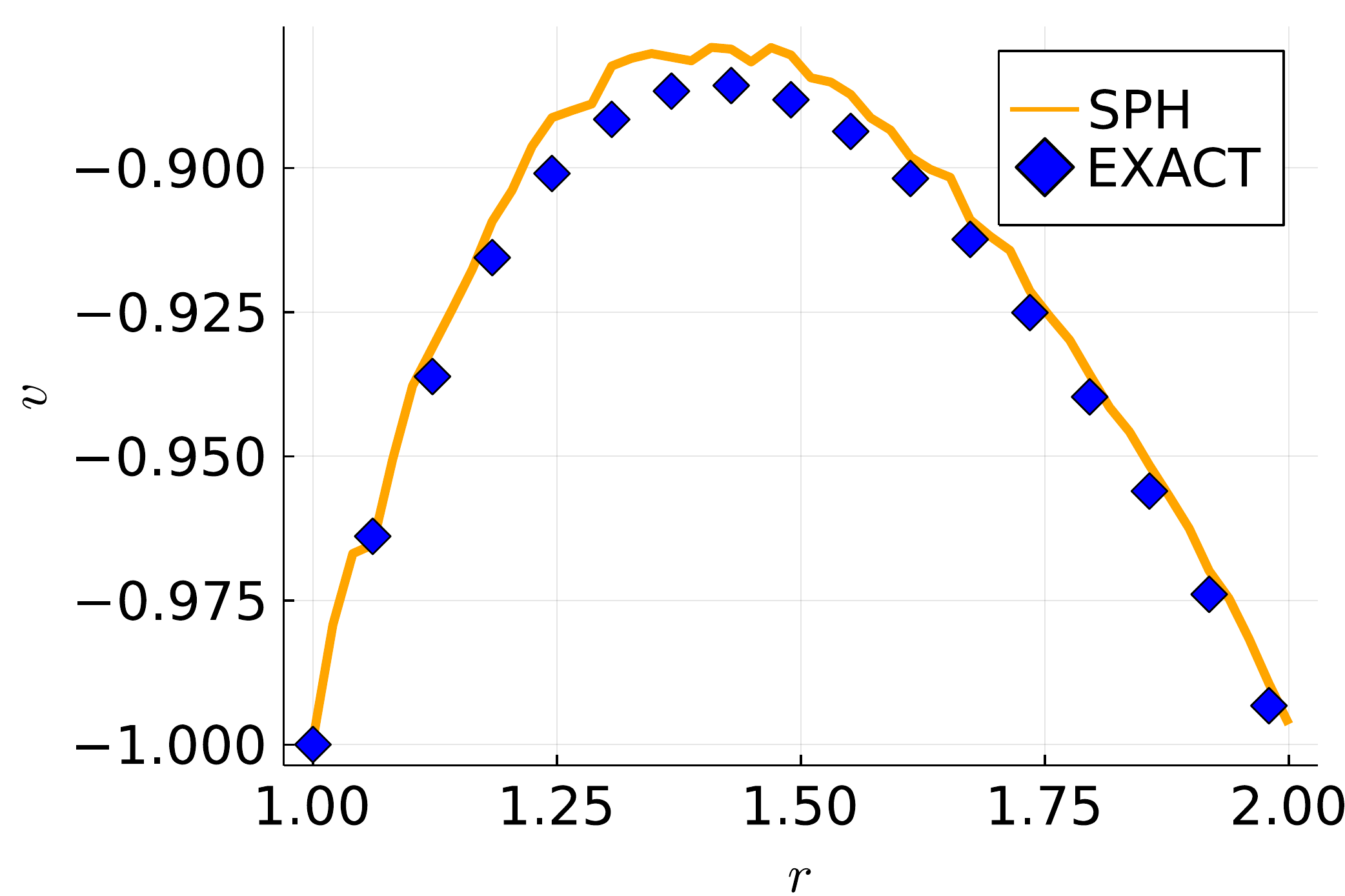}
		\caption{$v_y - \omega r$}
	\end{subfigure}
	\caption{Tangential velocity in the simulation at $t = 10$ along the segment $y = 0$, $R_1 \leq x \leq R_2$ (orange line) and its comparison to the exact solution (blue squares). Picture on the right magnifies error by subtracting a linear approximation $v_y \approx \omega r$ from both data arrays.}
	\label{fig:taco-speed}
\end{figure}
\begin{figure}[!htb]
	\begin{subfigure}{0.32\textwidth}
		\includegraphics[draft=false,width = \textwidth, trim={2cm 2cm 2cm 2cm}, 
		clip]{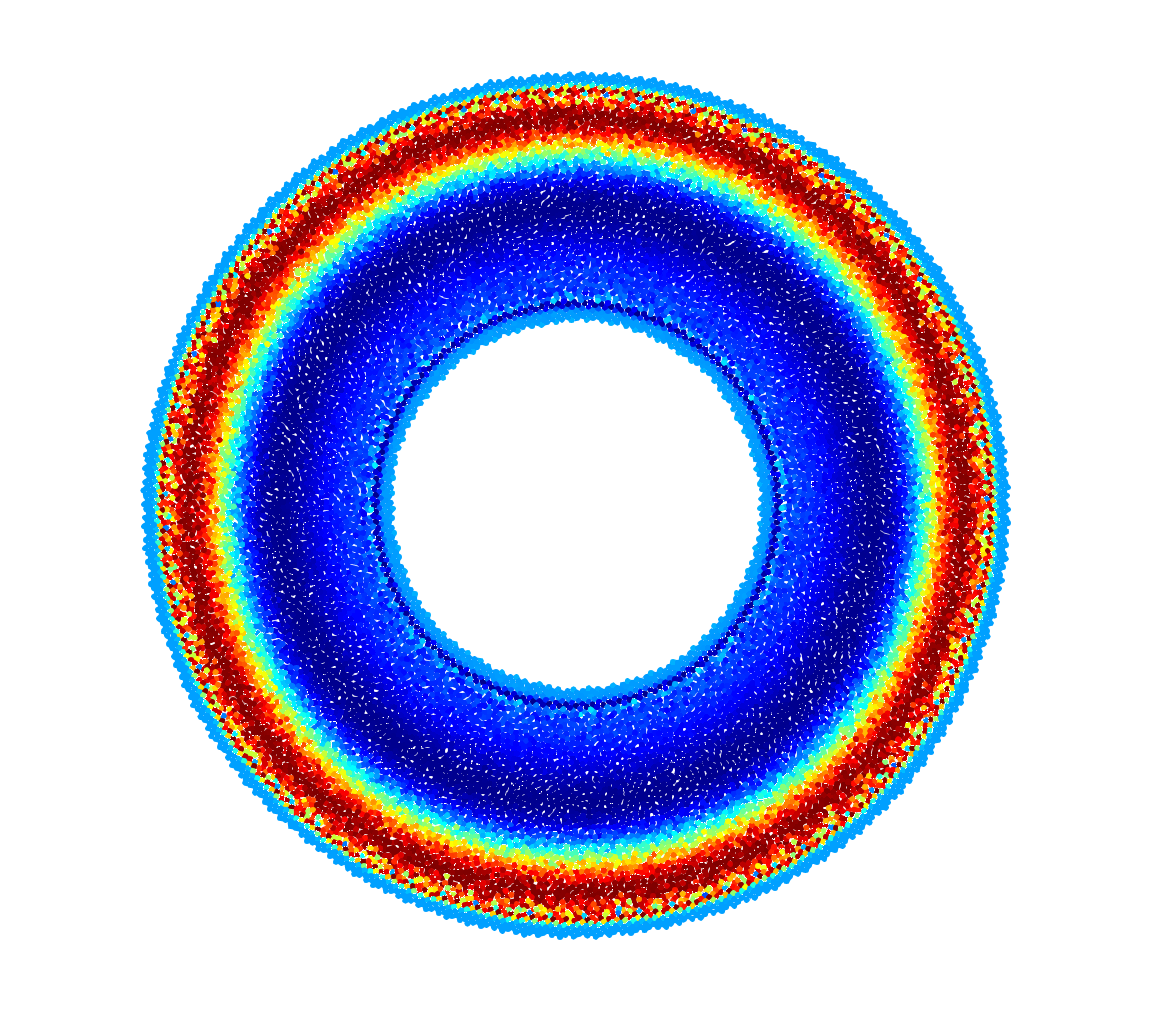}
		\caption{$t = 2$}
	\end{subfigure}
	\begin{subfigure}{0.32\textwidth}
		\includegraphics[draft=false,width = \textwidth, trim={2cm 2cm 2cm 2cm}, 
		clip]{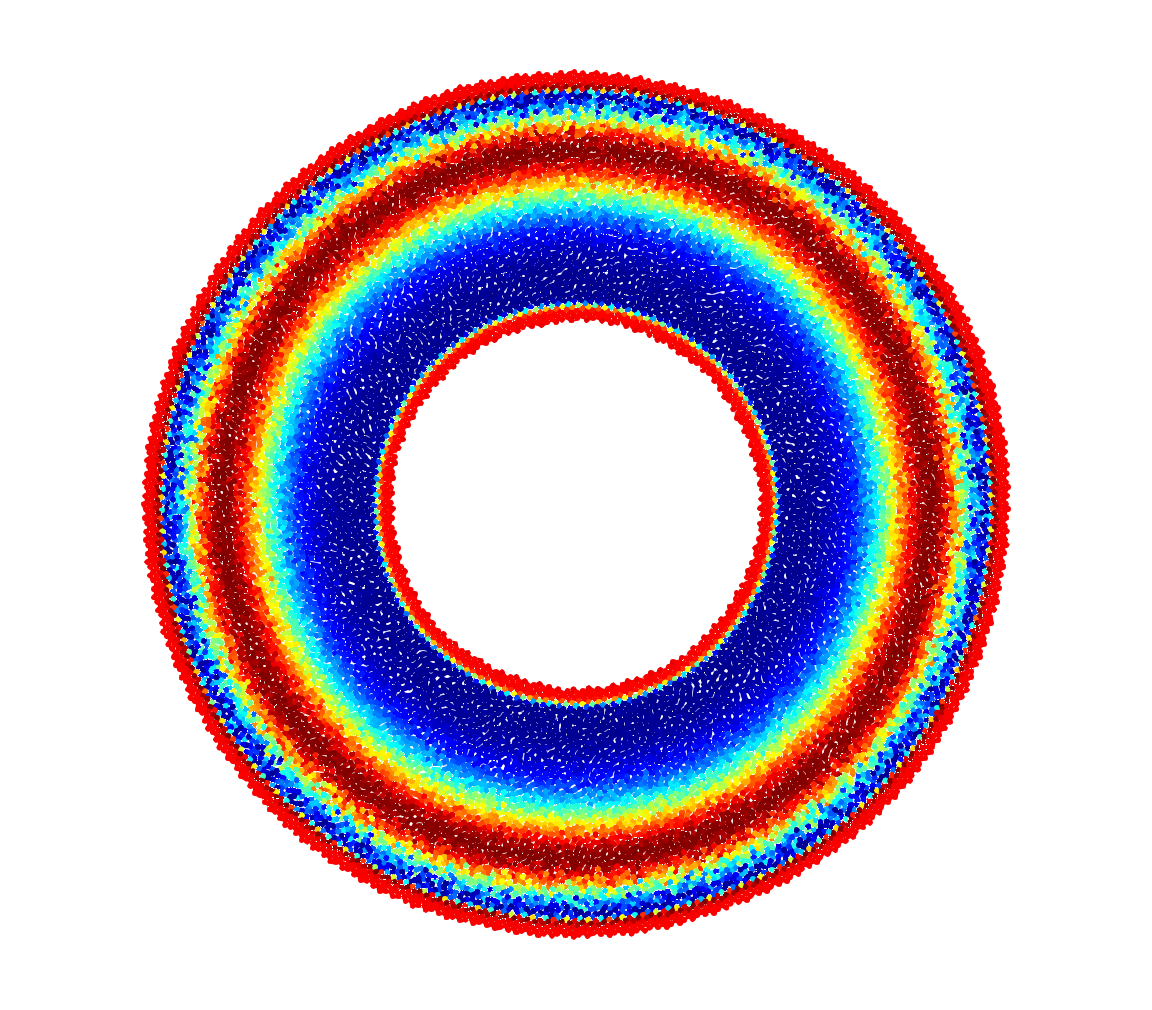}
		\caption{$t = 3$}
	\end{subfigure}
	\begin{subfigure}{0.32\textwidth}
		\includegraphics[draft=false,width = \textwidth, trim={2cm 2cm 2cm 2cm}, 
		clip]{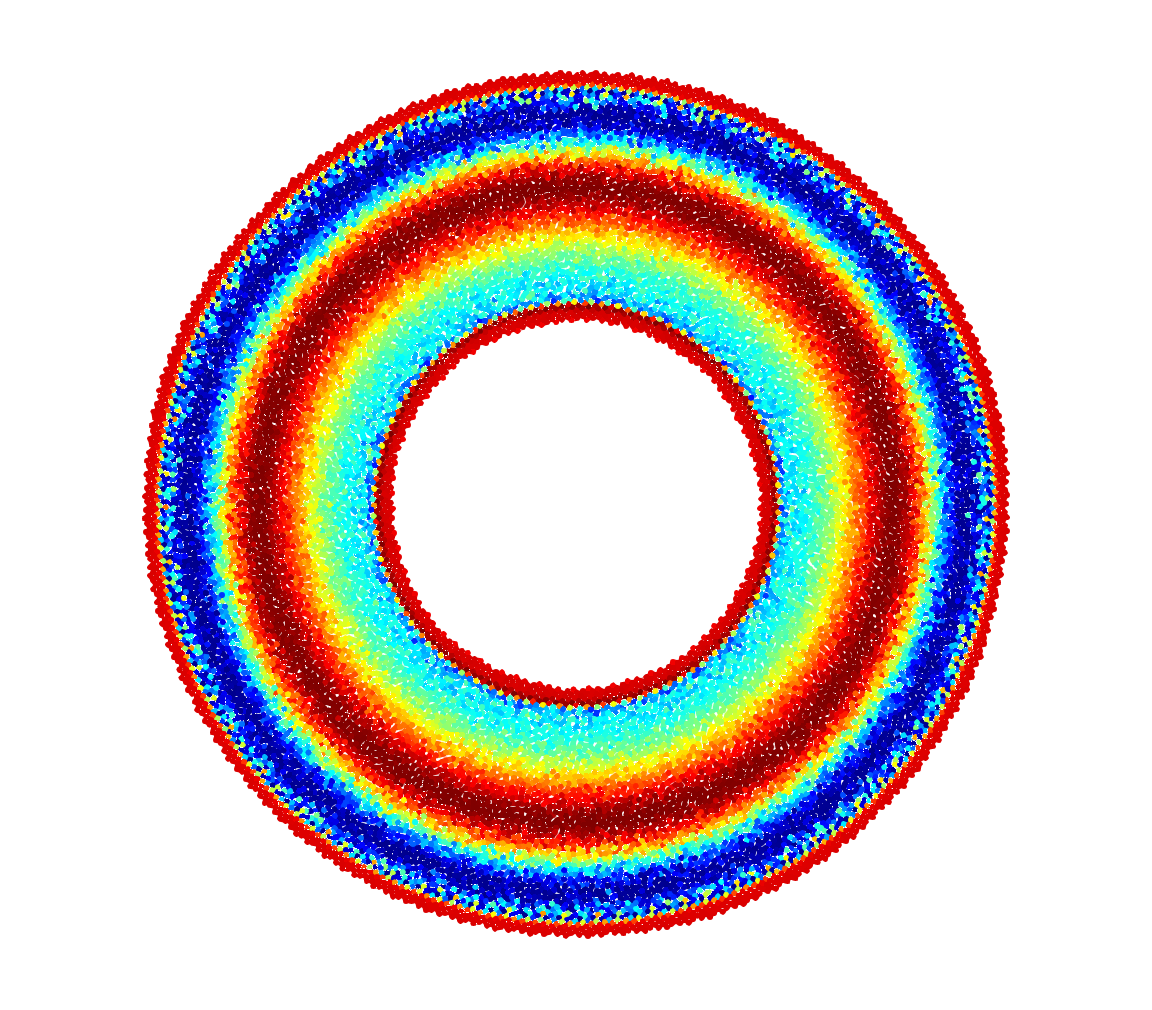}
		\caption{$t = 4$}
	\end{subfigure}
	\begin{subfigure}{0.32\textwidth}
		\includegraphics[draft=false,width = \textwidth, trim={2cm 2cm 2cm 2cm}, 
		clip]{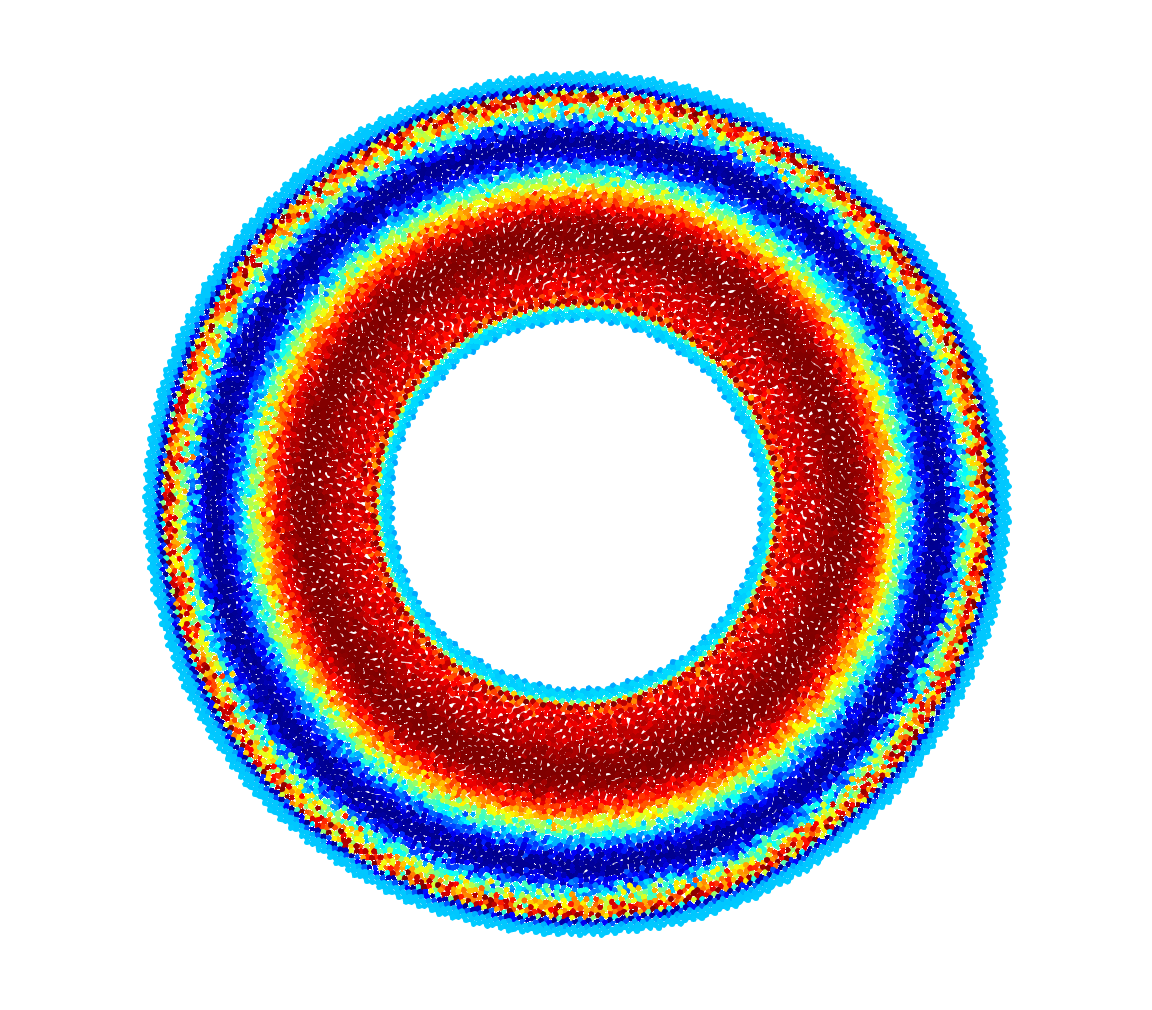}
		\caption{$t = 5$}
	\end{subfigure}
	\begin{subfigure}{0.32\textwidth}
		\includegraphics[draft=false,width = \textwidth, trim={2cm 2cm 2cm 2cm}, 
		clip]{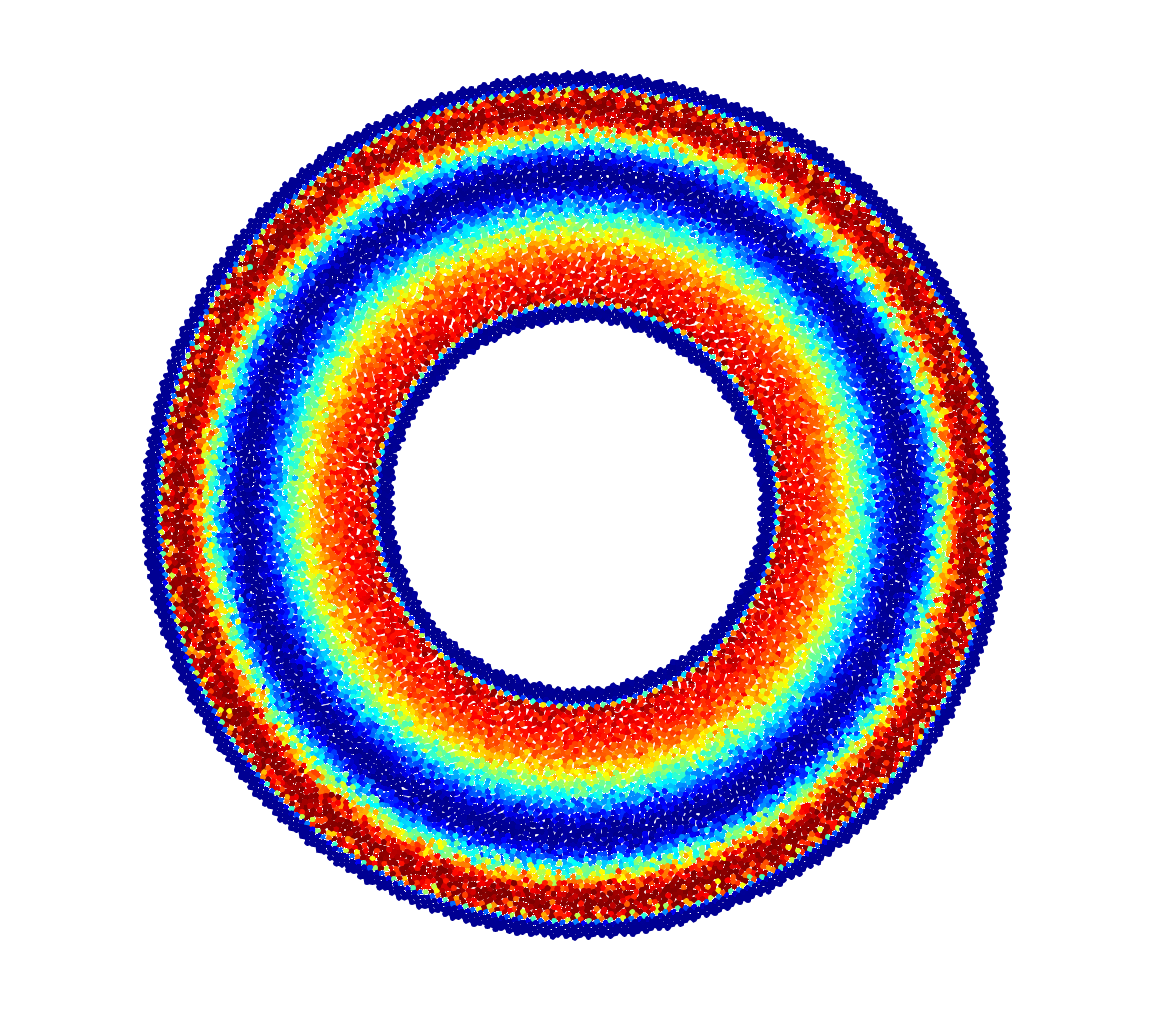}
		\caption{$t = 6$}
	\end{subfigure}
	\begin{subfigure}{0.32\textwidth}
		\includegraphics[draft=false,width = \textwidth, trim={2cm 2cm 2cm 2cm}, 
		clip]{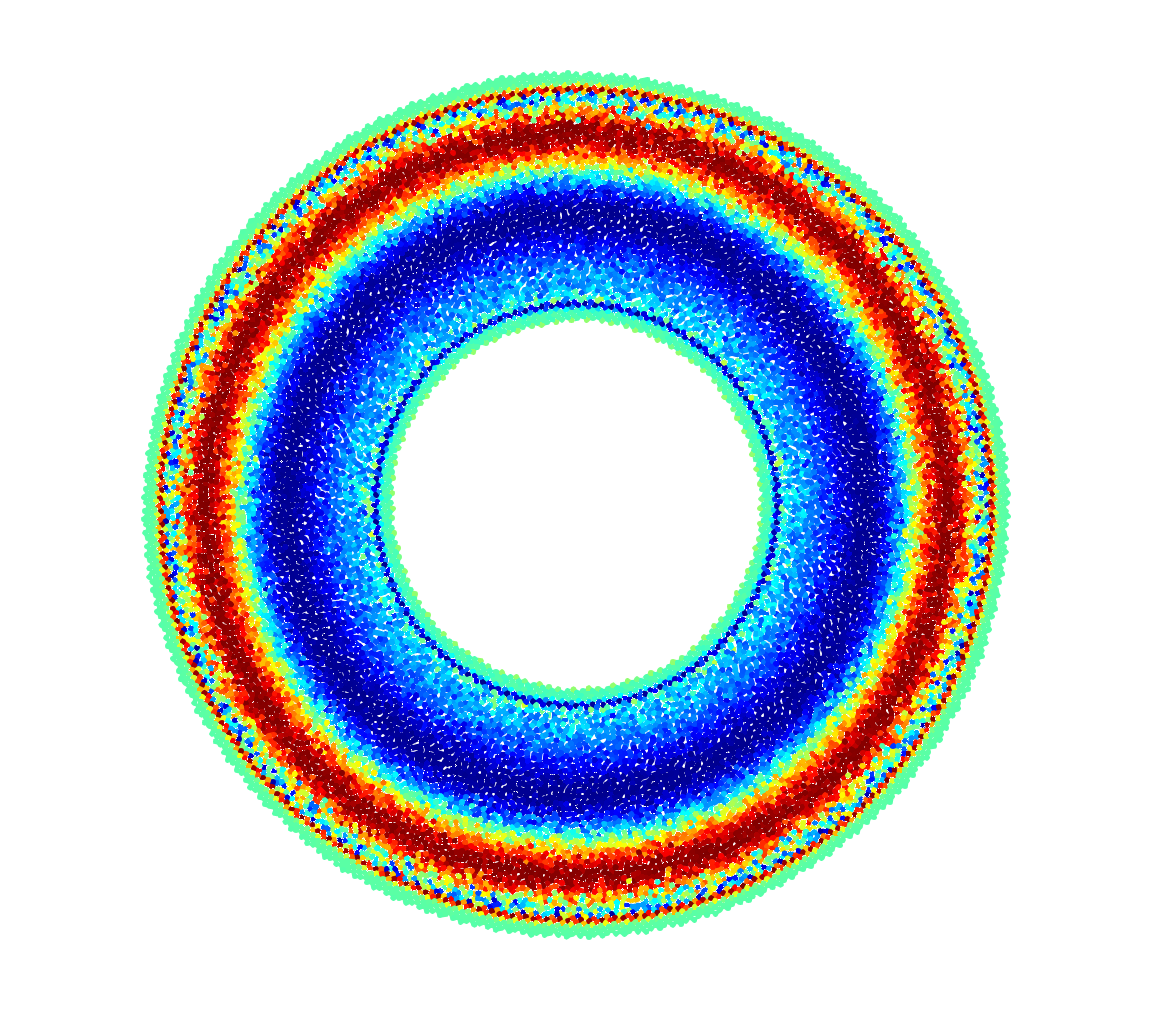}
		\caption{$t = 7$}
	\end{subfigure}
	\centering
	\includegraphics[draft=false,width= 0.5 \textwidth]{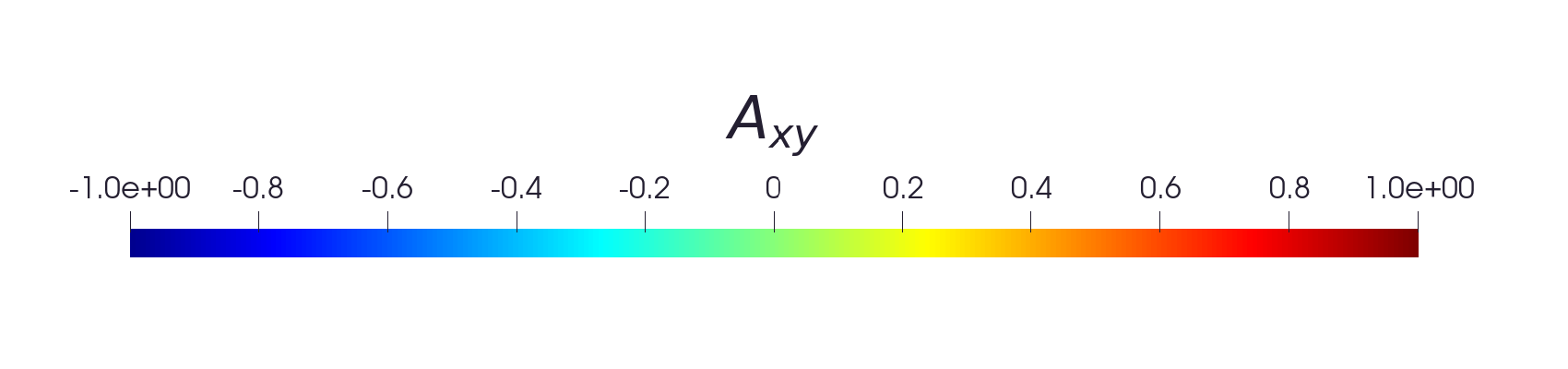}
	\caption{Plot of $xy$ distortion component. Interestingly, whereas velocity converges in time (approximately) to the stationary solution \eqref{eq:taco-exact}, the $\bbA$ field is always changing in a periodic manner, forming waves that travel from the outer ring towards the center.}
	\label{fig:taco-pics}
\end{figure}
We now turn our attention to the fluid regime of the SHTC equations, which means that $\tau < 
\infty$ and the relaxation terms need to be taken into account. First, we try a simple test of 
laminar flow 
in an annulus
$$ \Omega = \{ (x,y) : R_1^2 < x^1 + y^2 < R_2^2\} $$
driven by a rigid counterclockwise rotation of the outer ring with angular velocity $\omega$. 
Meanwhile, the inner ring has zero angular velocity. We take $R_1 = 1$, $R_2 = 2$ and $\omega = 1$. 
The flow is incompressible and well described by the incompressible Navier-Stokes equations with 
kinetic viscosity $\nu = 0.1$. This corresponds to a low Reynolds number
$$ \text{Re} = \frac{\omega R_2 (R_2 - R_1)}{\nu} = 20,$$
which ensures laminar flow. The exact stationary solution to this problem is given by the formula
\begin{equation}
	\vv = \frac{R_2}{r} \frac{\frac{r}{R_1} - \frac{R_1}{r}}{\frac{R_2}{R_1} - \frac{R_1}{R_2}} \begin{pmatrix*}[r]
		-\omega y\\
		\omega x
	\end{pmatrix*}.
	\label{eq:taco-exact}
\end{equation}
The solution theoretically does not depend on $\nu$ but viscosity affects how quickly the velocity field converges (if at all), starting from $\vv = \bm{0}$.

The Navier-Stokes equations (NSEs) are formally incompatible with the SHTC equations, which can be 
inferred from the fact that NSEs are a hyperbolic-parabolic system, whereas SHTC equations include 
only first-order hyperbolic equations. However, it is possible to obtain NSEs (at least formally) 
in the asymptotic expansion of the SHTC equations as the first-order terms in $ \tau $ 
\cite{DPRZ2016}.  To achieve 
sufficiently 
small values of $ \tau $ one needs to take sufficiently large values of the shear sound speed since 
they are related as \cite{DPRZ2016}
\begin{equation}
	\tau = \frac{6\nu}{c_s^2}.
	\label{eq:tau-from-nu}
\end{equation} 
Here, we essentially mirror the common approach in SPH, 
where incompressibility is enforced by ``sufficiently high'' values of $c_0$ (corresponding to 
small Mach number $ 
\mathrm{Ma} = \Vert\vv\Vert/c_0 \ll 1 $) but this time for the shear 
component of energy. The characteristic speed in this simulation is $\omega R_2 = 2$, so it is 
reasonable to take $c_0 = 20$ and $c_s = 40$. Here, we set $\delta r = \frac{1}{40}$ and $\delta t$ 
according to \eqref{eq:time-step-size}. With these parameters, we get 
$$ \frac{\delta t}{\tau} \doteq 0.075$$
so the natural time step is significantly smaller than $\tau$, justifying the use of an explicit time integrator for relaxation.

Despite the numerous approximations used, we obtain reasonable agreement with the exact solution, 
as shown in Figure \ref{fig:taco-speed}. It is interesting here to plot the distortion field 
(Figure \ref{fig:taco-pics}). Even for such a simple \emph{stationary} flow, $\bbA$ displays 
non-stationary behavior 
due to the rotation of the local basis vectors represented by $ \bbA $, e.g. see 
\cite{nonNewtonian2021,DPRZ2016,HPR2016}. 
\subsection{Lid-driven cavity (Newtonian fluid)}
\begin{figure}[!htb]
	\begin{subfigure}{0.32\textwidth}
		\includegraphics[draft=false,width = \textwidth, trim={16cm 2cm 16cm 2cm}, 
		clip]{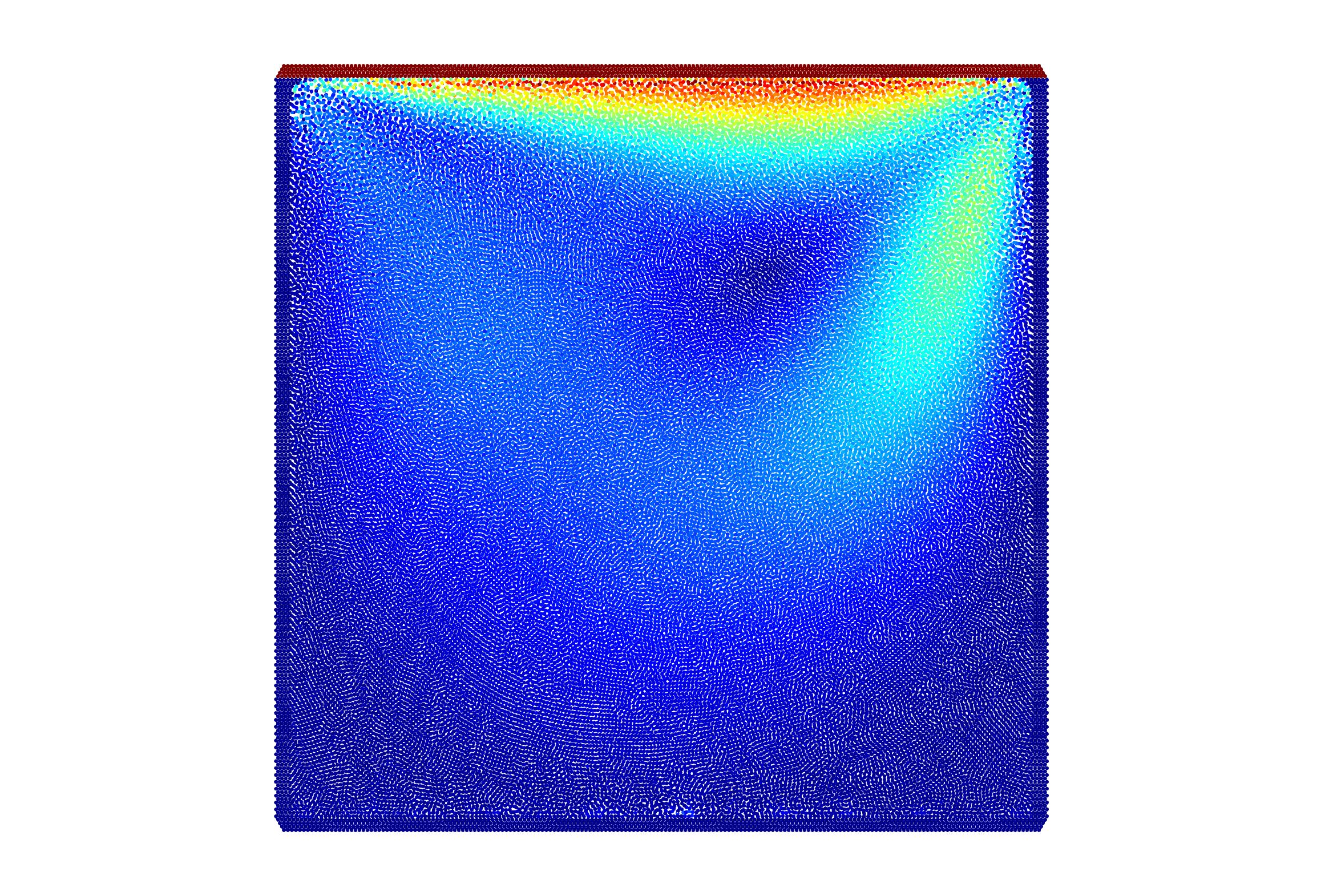}
		\caption{particles, $\mathrm{Re} = 100$, $t = 10$}
	\end{subfigure}
	\begin{subfigure}{0.32\textwidth}
		\includegraphics[draft=false,width = \textwidth, trim={16cm 2cm 16cm 2cm}, 
		clip]{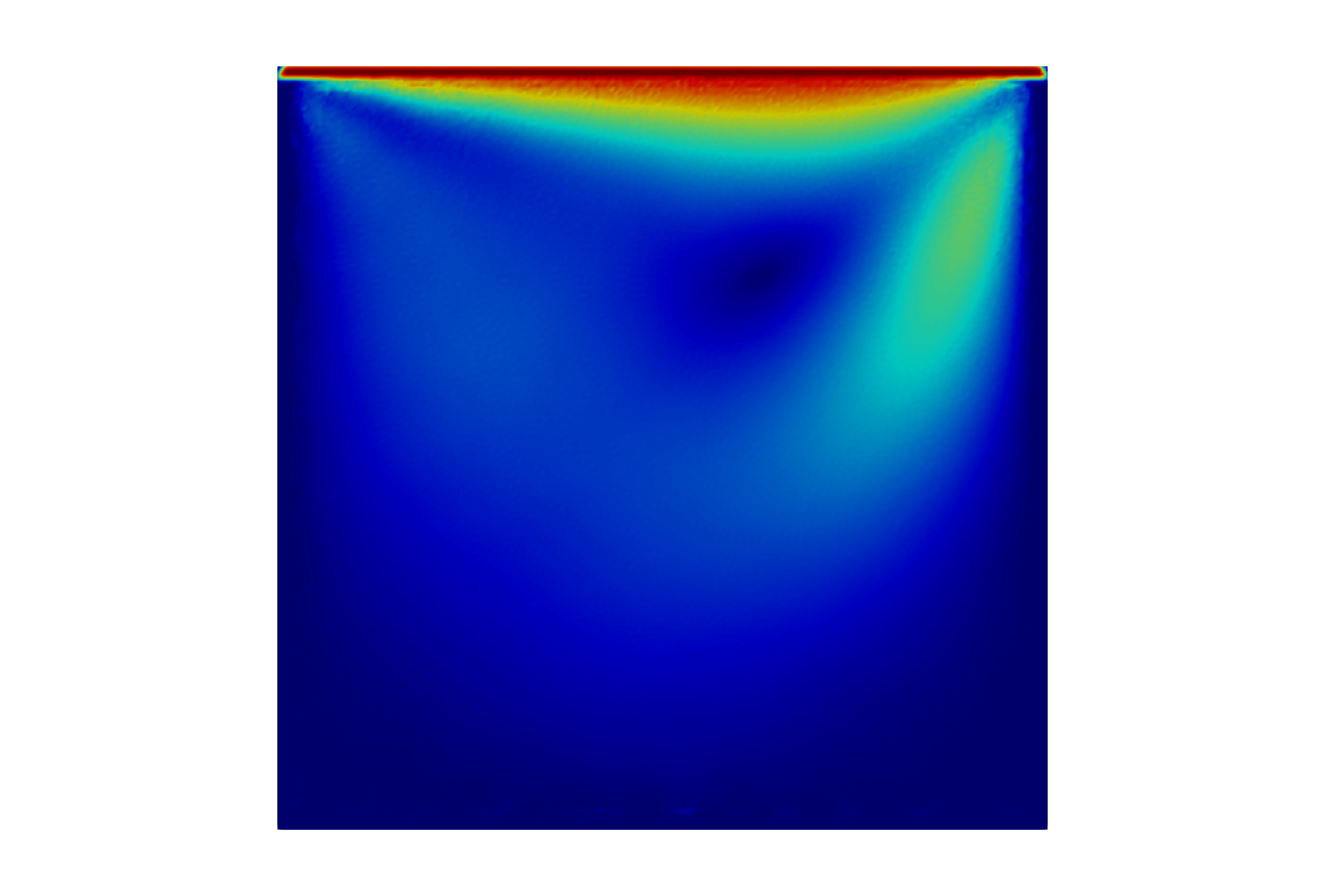}
		\caption{interpolation, $\mathrm{Re} = 100$, $t = 10$}
	\end{subfigure}
	\begin{subfigure}{0.32\textwidth}
		\includegraphics[draft=false,width = \textwidth, trim={16cm 2cm 16cm 2cm}, 
		clip]{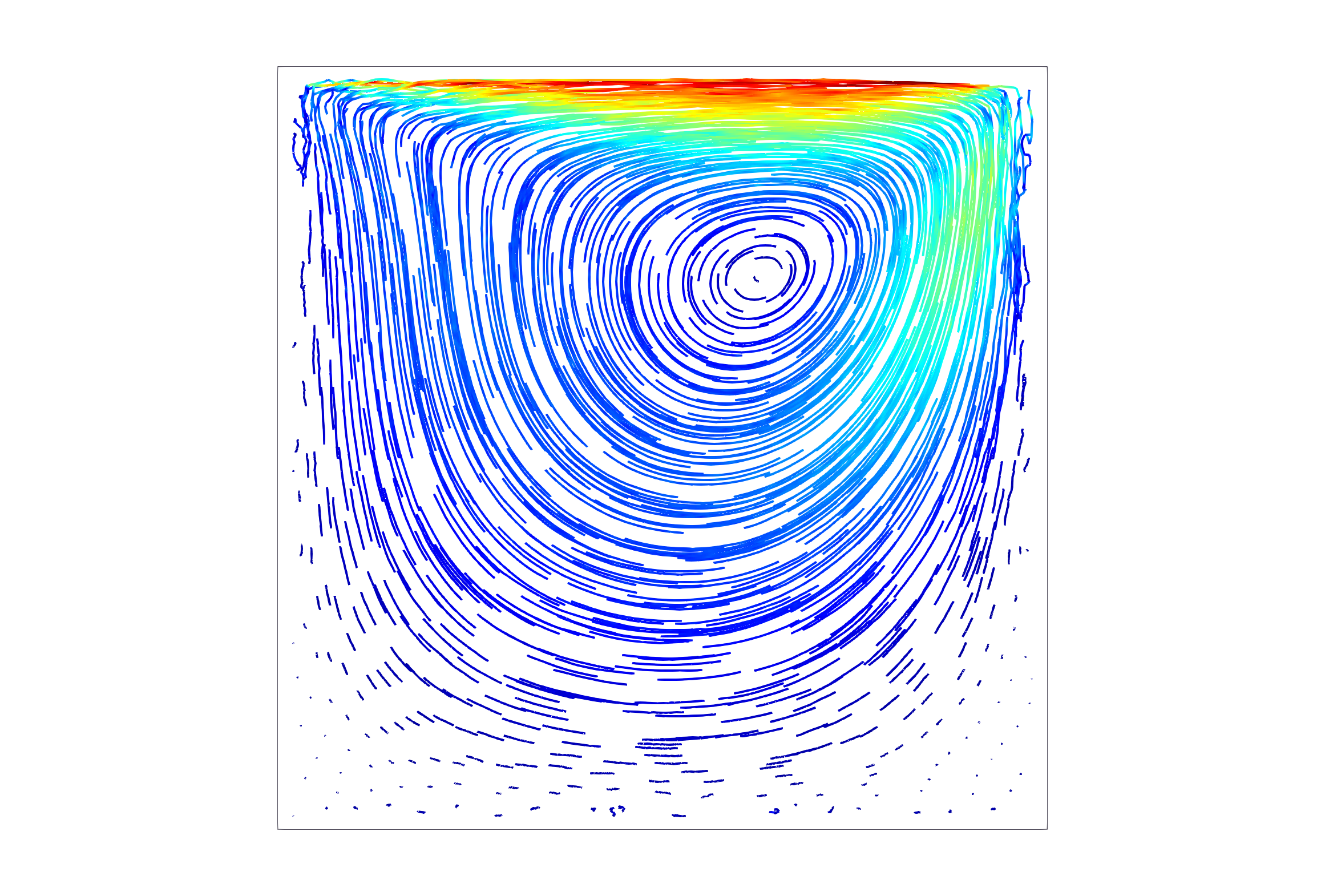}
		\caption{path-lines, $\mathrm{Re} = 100$, $t = 10$}
	\end{subfigure}
	
	\begin{subfigure}{0.32\textwidth}
		\includegraphics[draft=false,width = \textwidth, trim={16cm 2cm 16cm 2cm}, 
		clip]{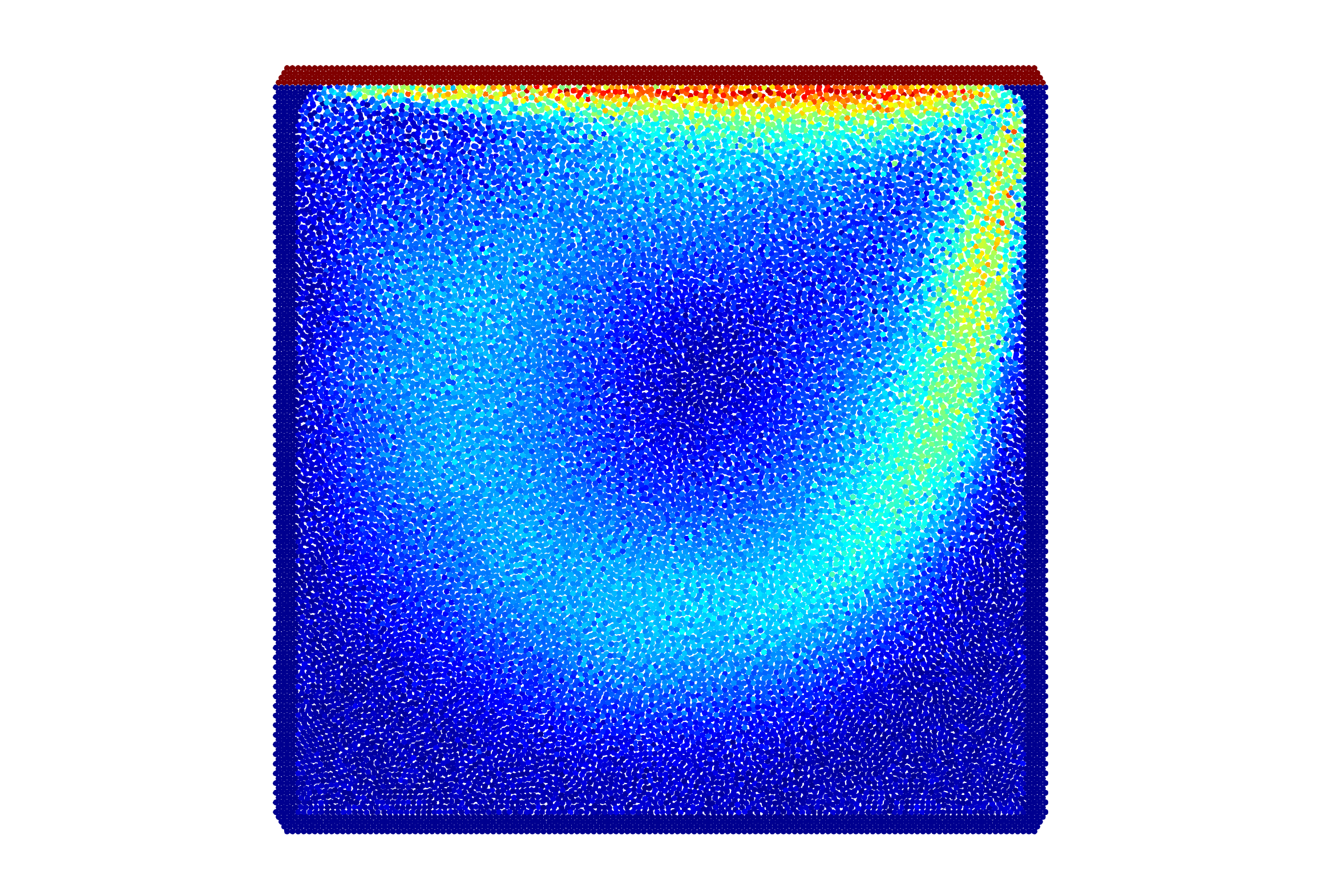}
		\caption{particles, $\mathrm{Re} = 400$, $t = 20$}
	\end{subfigure}
	\begin{subfigure}{0.32\textwidth}
		\includegraphics[draft=false,width = \textwidth, trim={16cm 2cm 16cm 2cm}, 
		clip]{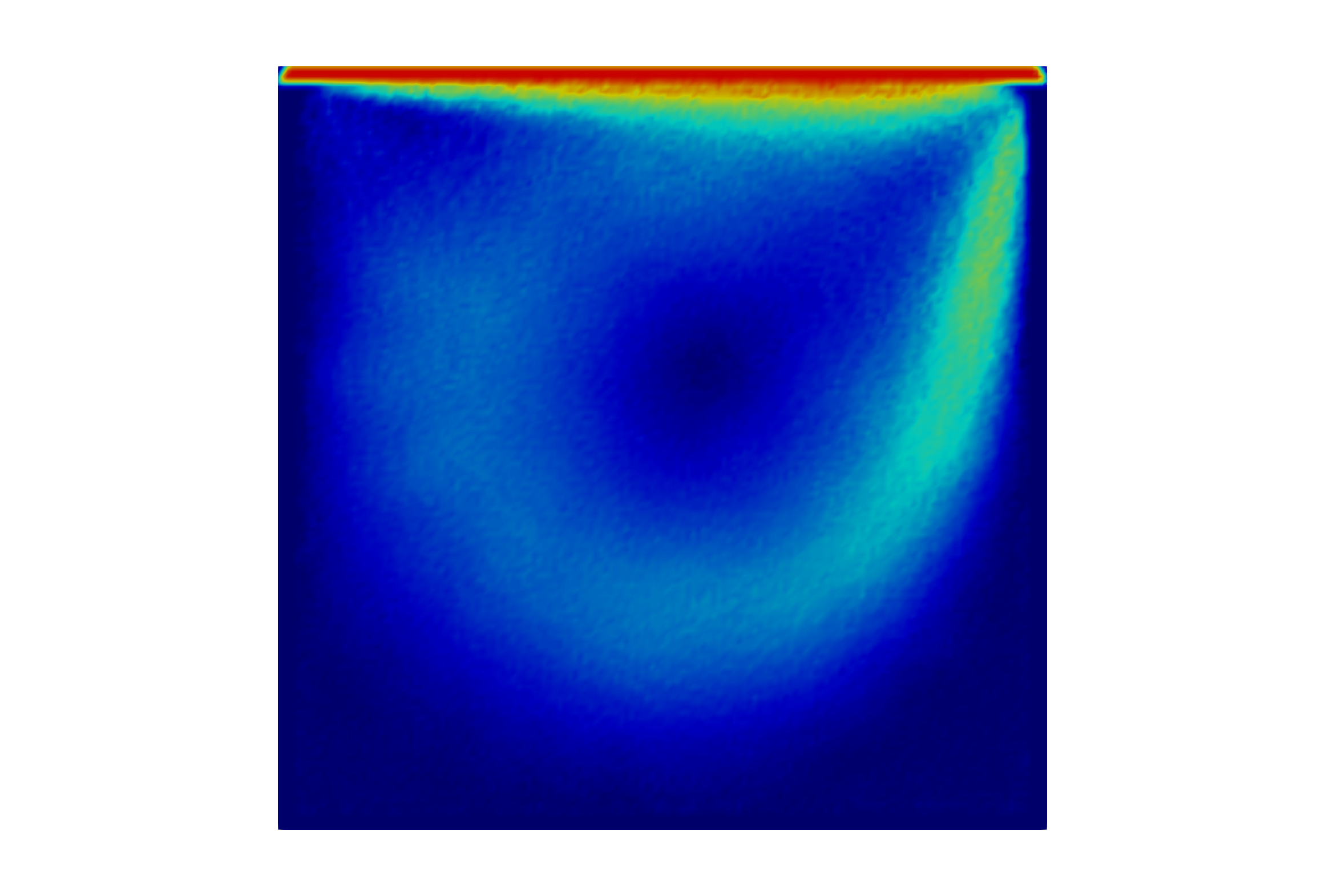}
		\caption{interpolation, $\mathrm{Re} = 400$, $t = 20$}
	\end{subfigure}
	\begin{subfigure}{0.32\textwidth}
		\includegraphics[draft=false,width = \textwidth, trim={16cm 2cm 16cm 2cm}, 
		clip]{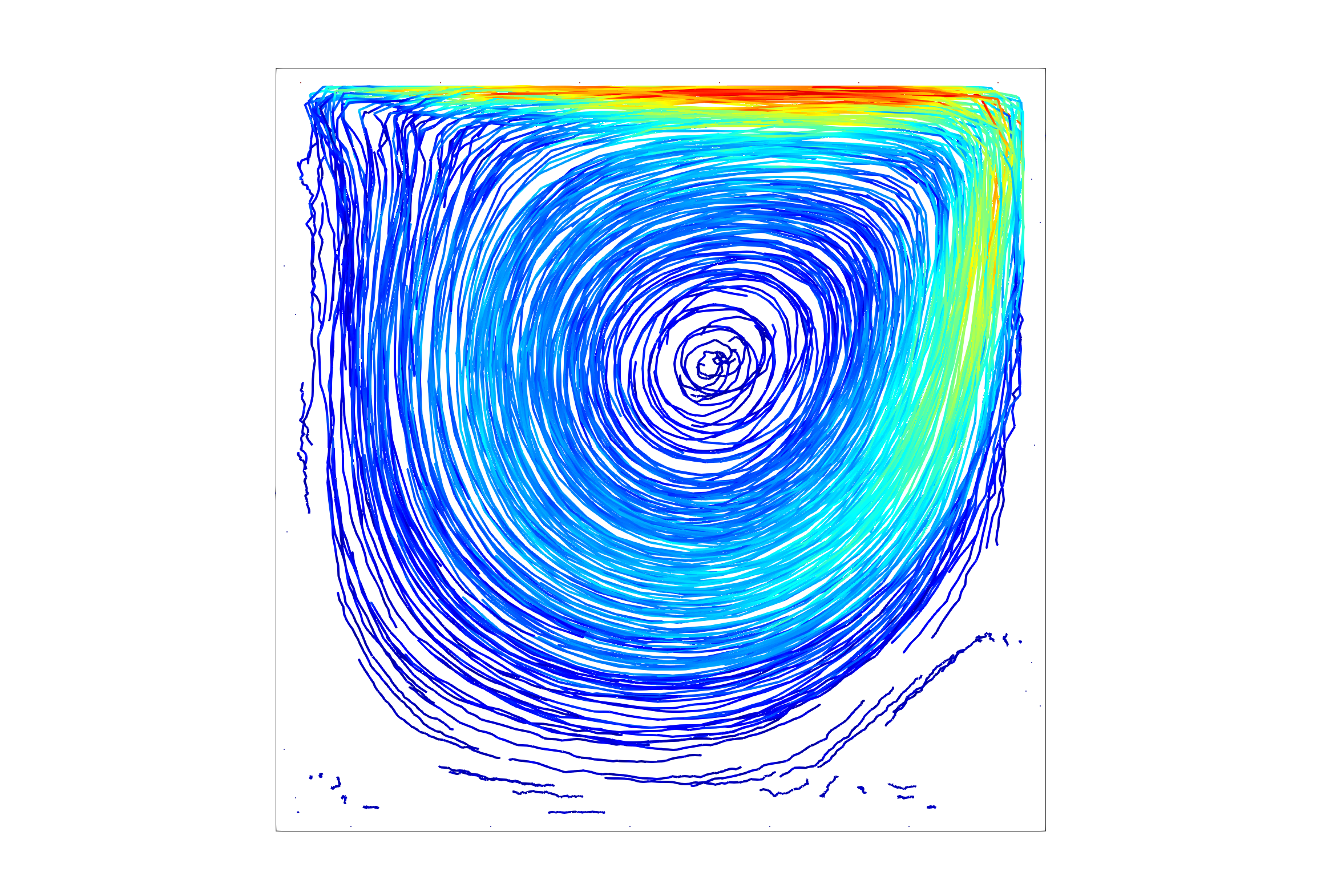}
		\caption{path-lines, $\mathrm{Re} = 400$, $t = 20$}
	\end{subfigure}
	\centering
	\includegraphics[draft=false,width= 0.5 \textwidth]{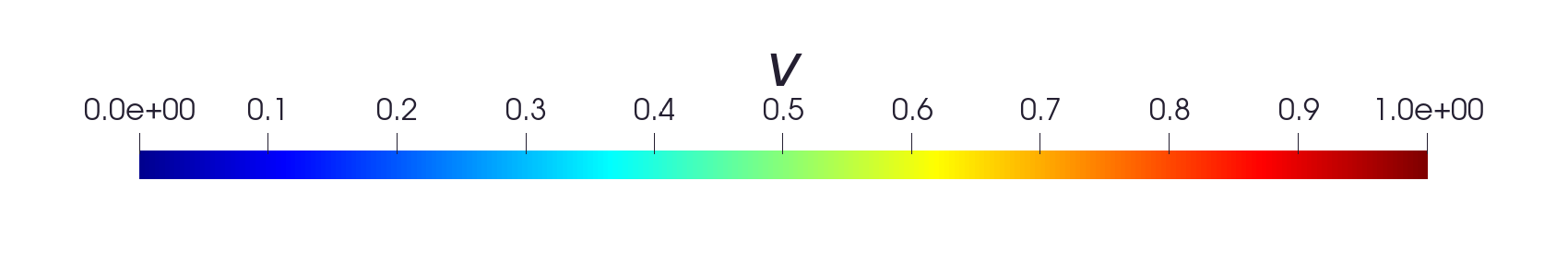}
	\caption{Velocity field in the Newtonian lid-driven cavity benchmark.}
	\label{fig:ldc-v}
\end{figure}
\begin{figure}[!htb]
	\begin{subfigure}{0.32\textwidth}
		\includegraphics[draft=false,width = \textwidth, trim={16cm 2cm 16cm 2cm}, 
		clip]{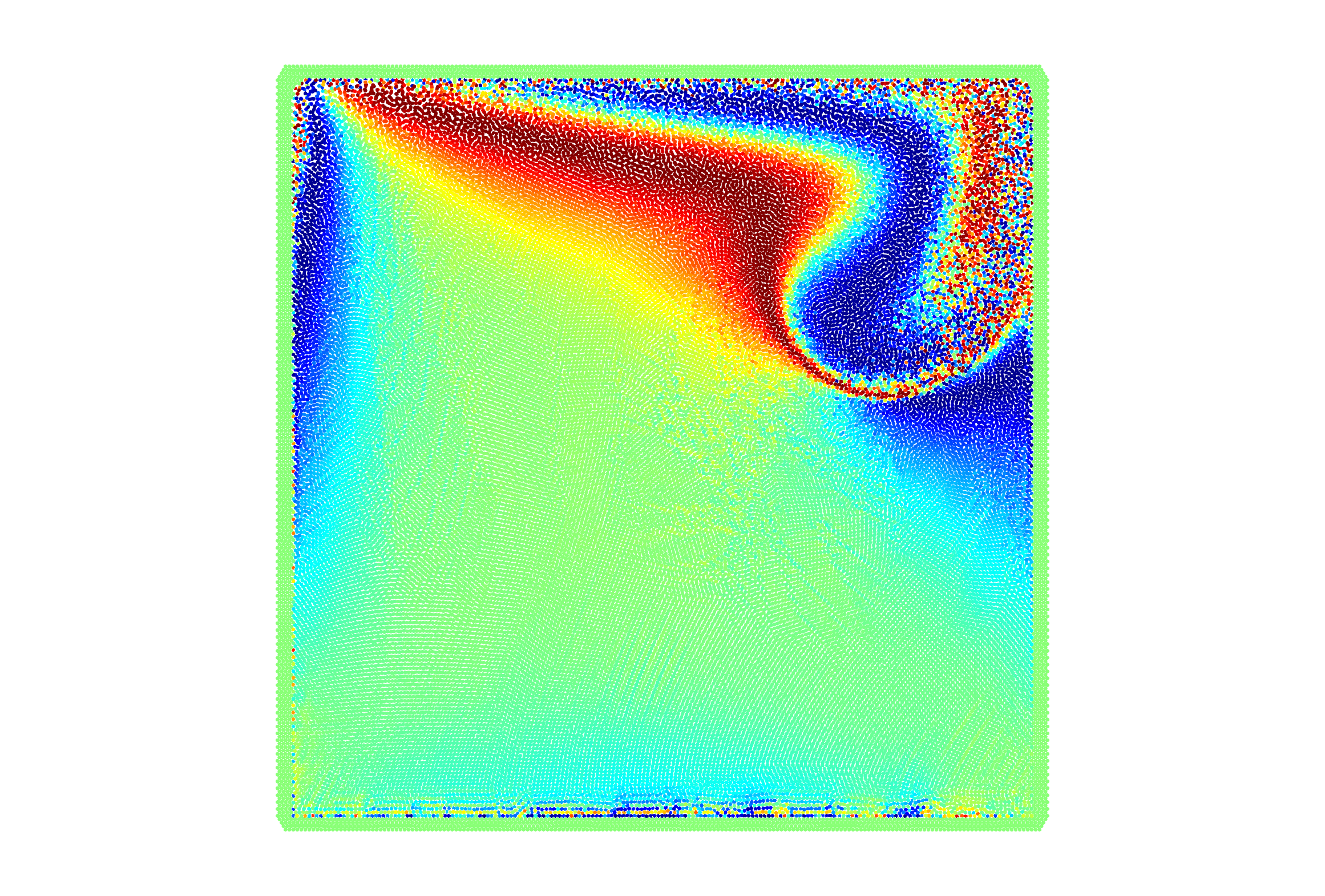}
		\caption{$t = 2$}
	\end{subfigure}
	\begin{subfigure}{0.32\textwidth}
		\includegraphics[draft=false,width = \textwidth, trim={16cm 2cm 16cm 2cm}, 
		clip]{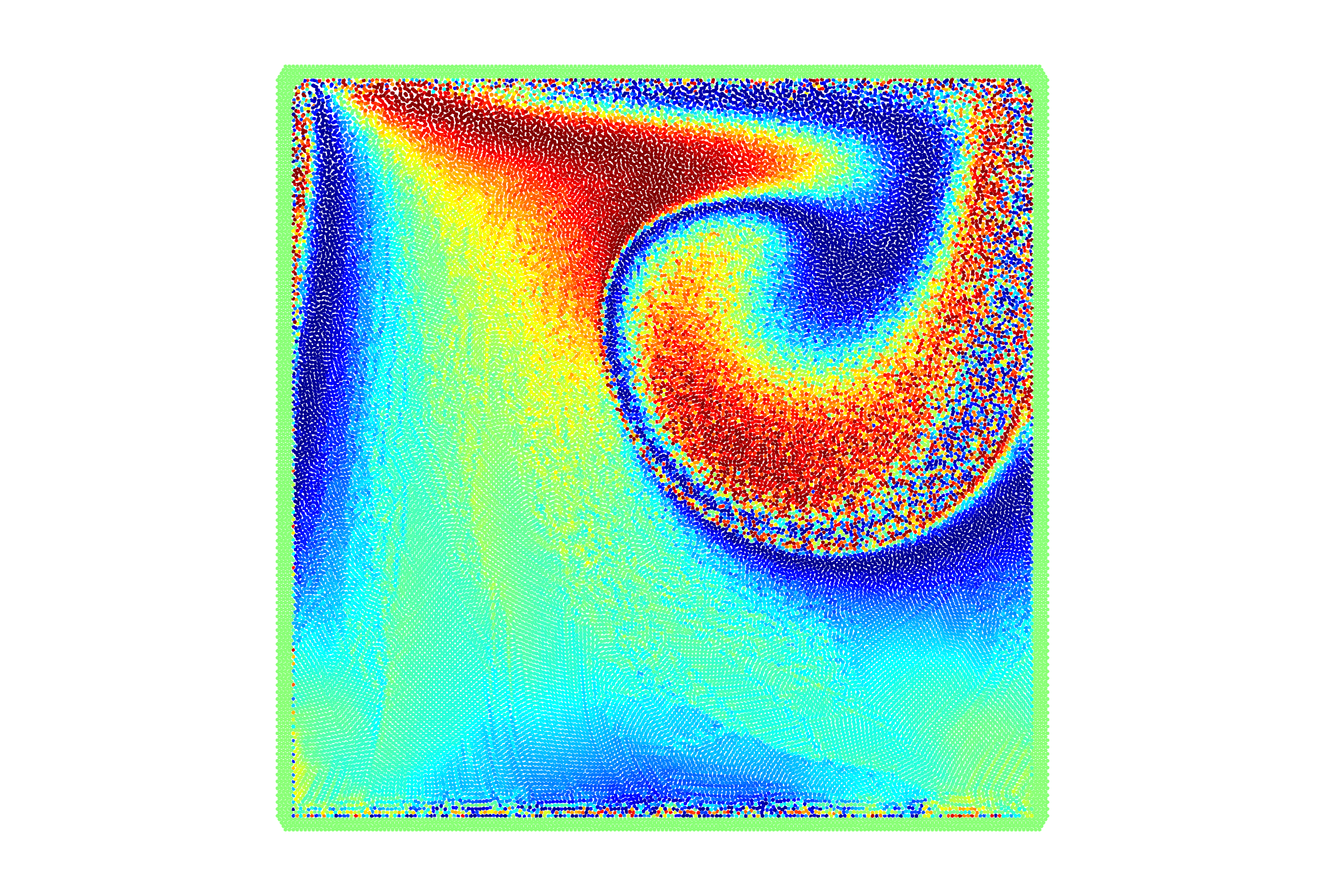}
		\caption{$t = 4$}
	\end{subfigure}
	\begin{subfigure}{0.32\textwidth}
		\includegraphics[draft=false,width = \textwidth, trim={16cm 2cm 16cm 2cm}, 
		clip]{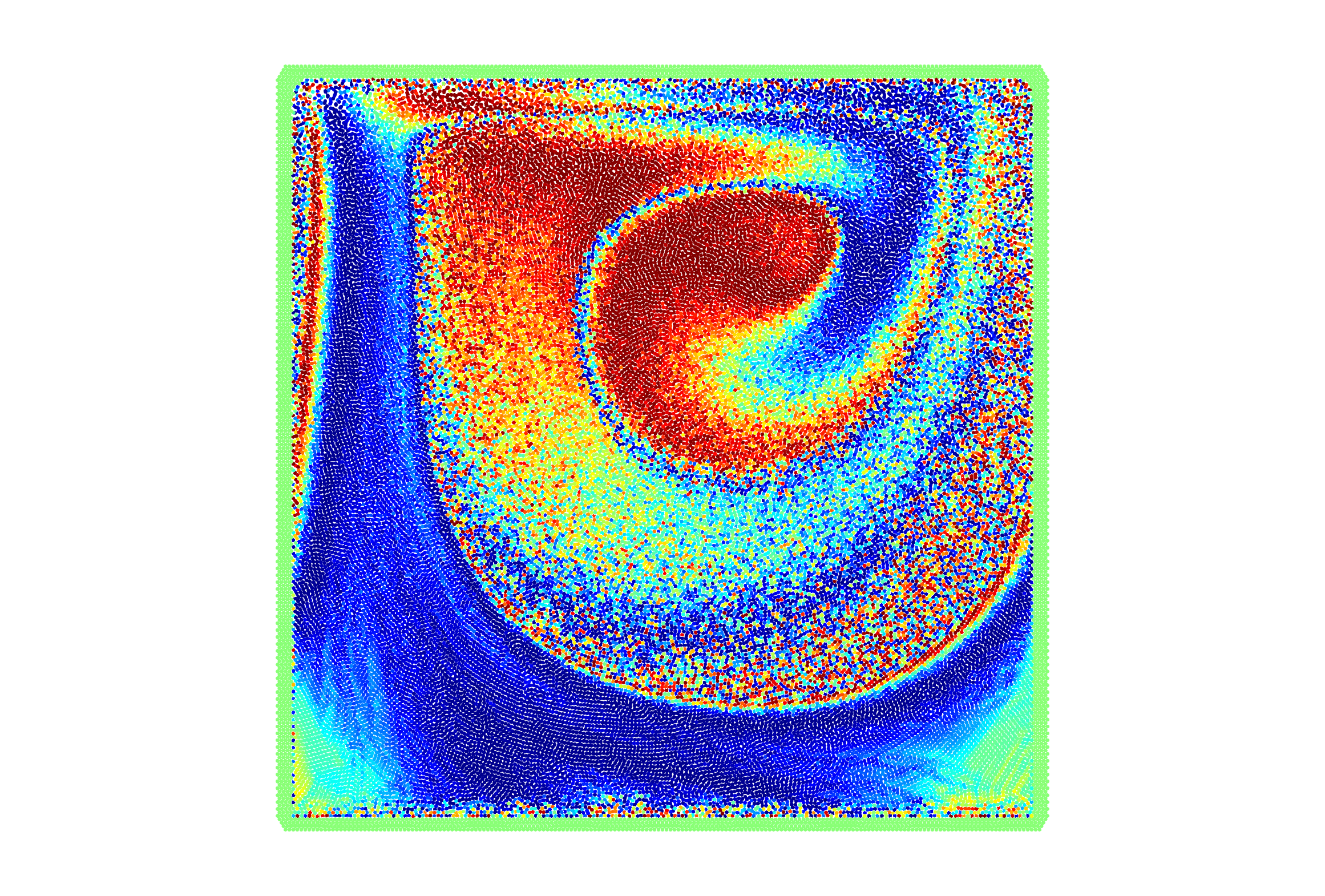}
		\caption{$t = 10$}
	\end{subfigure}
	\centering
	\includegraphics[draft=false,width= 0.5 \textwidth]{images/ldc-colorbar-A.png}
	\caption{Evolution of distortion field in Newtonian lid-driven cavity benchmark for $\mathrm{Re} = 100$.}
	\label{fig:ldc-A}
\end{figure}
\begin{figure}
	\centering		
	\begin{subfigure}{0.49\textwidth}
		\includegraphics[draft=false,width = \textwidth]{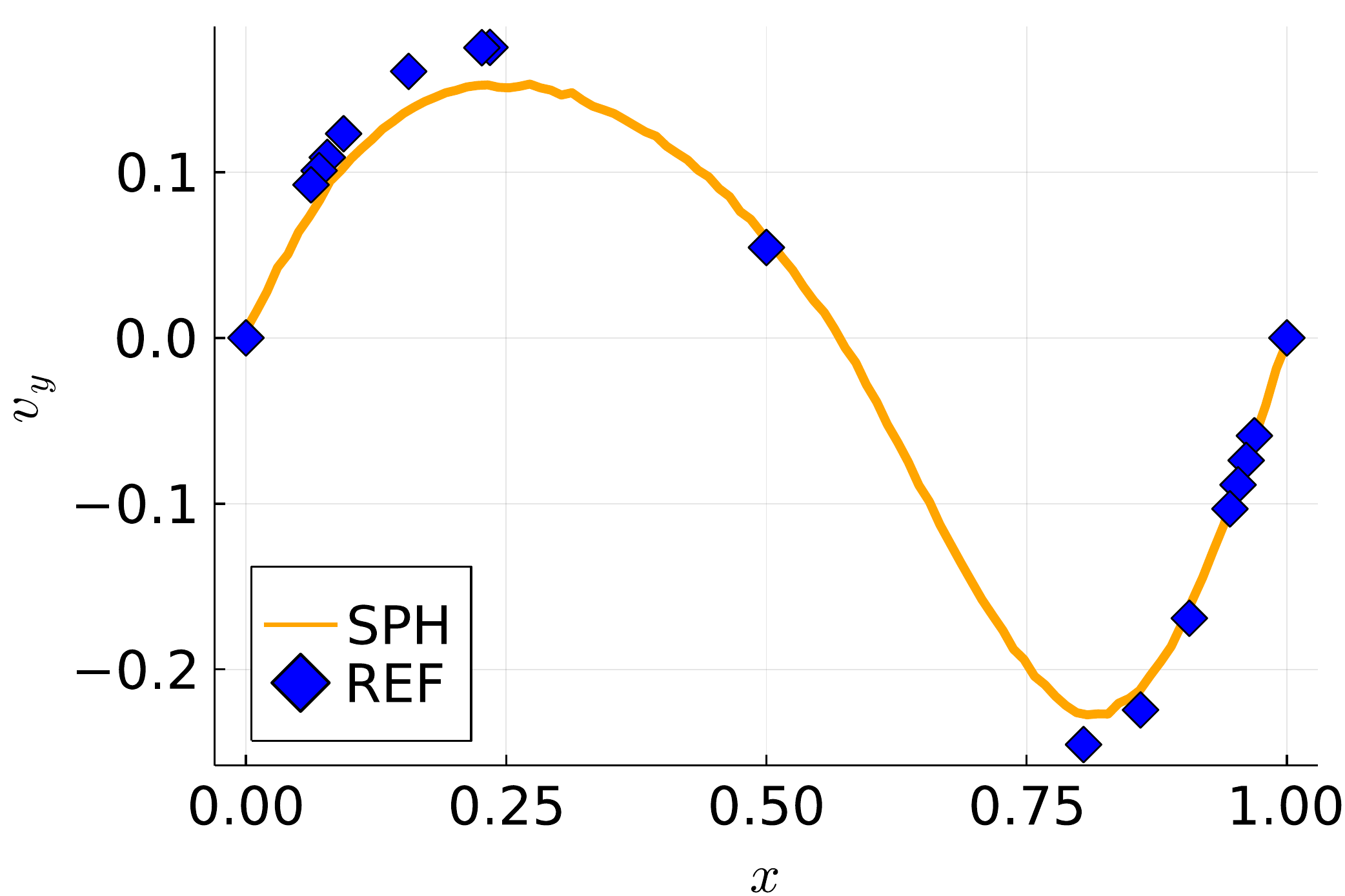}
		\caption{$v_y$ along $x$-centerline, $\mathrm{Re} = 100$, $t = 10$}
	\end{subfigure}
	\begin{subfigure}{0.49\textwidth}
		\includegraphics[draft=false,width = \textwidth]{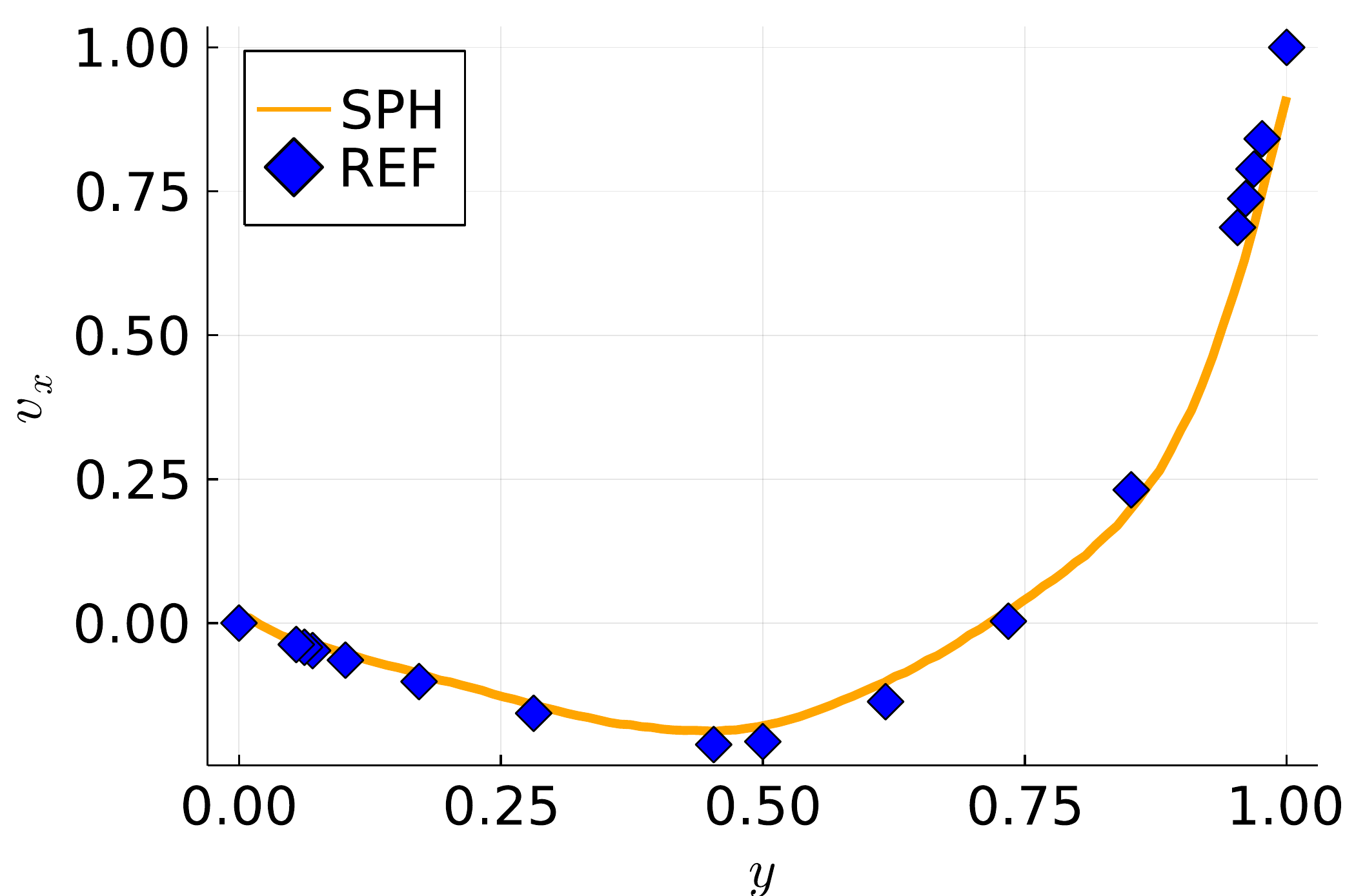}
		\caption{$v_x$ along $y$-centerline, $\mathrm{Re} = 100$, $t = 10$}
	\end{subfigure}
	\begin{subfigure}{0.49\textwidth}
		\includegraphics[draft=false,width = \textwidth]{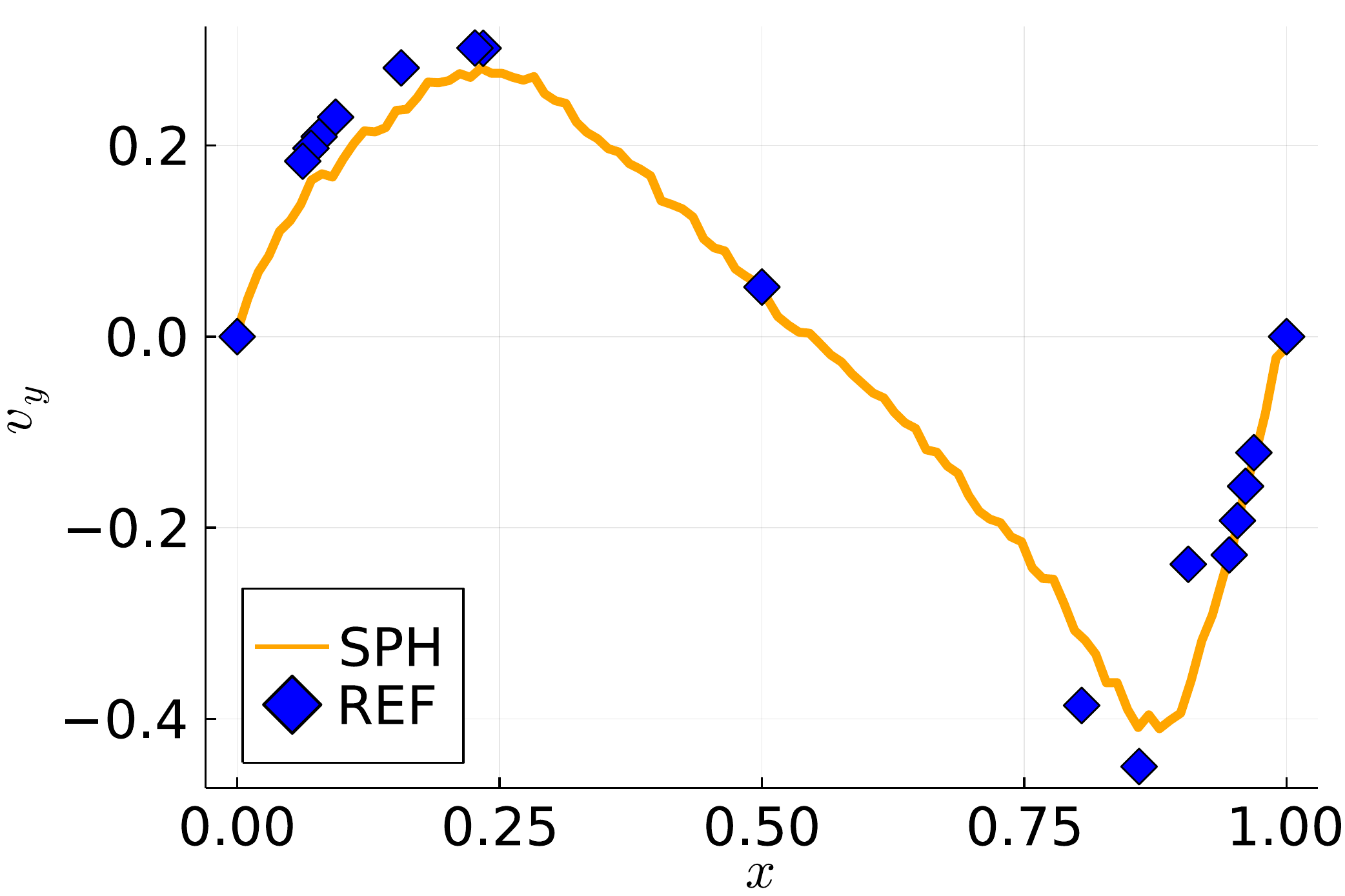}
		\caption{$v_y$ along $x$-centerline, $\mathrm{Re} = 400$, $t = 10$}
	\end{subfigure}
	\begin{subfigure}{0.49\textwidth}
		\includegraphics[draft=false,width = \textwidth]{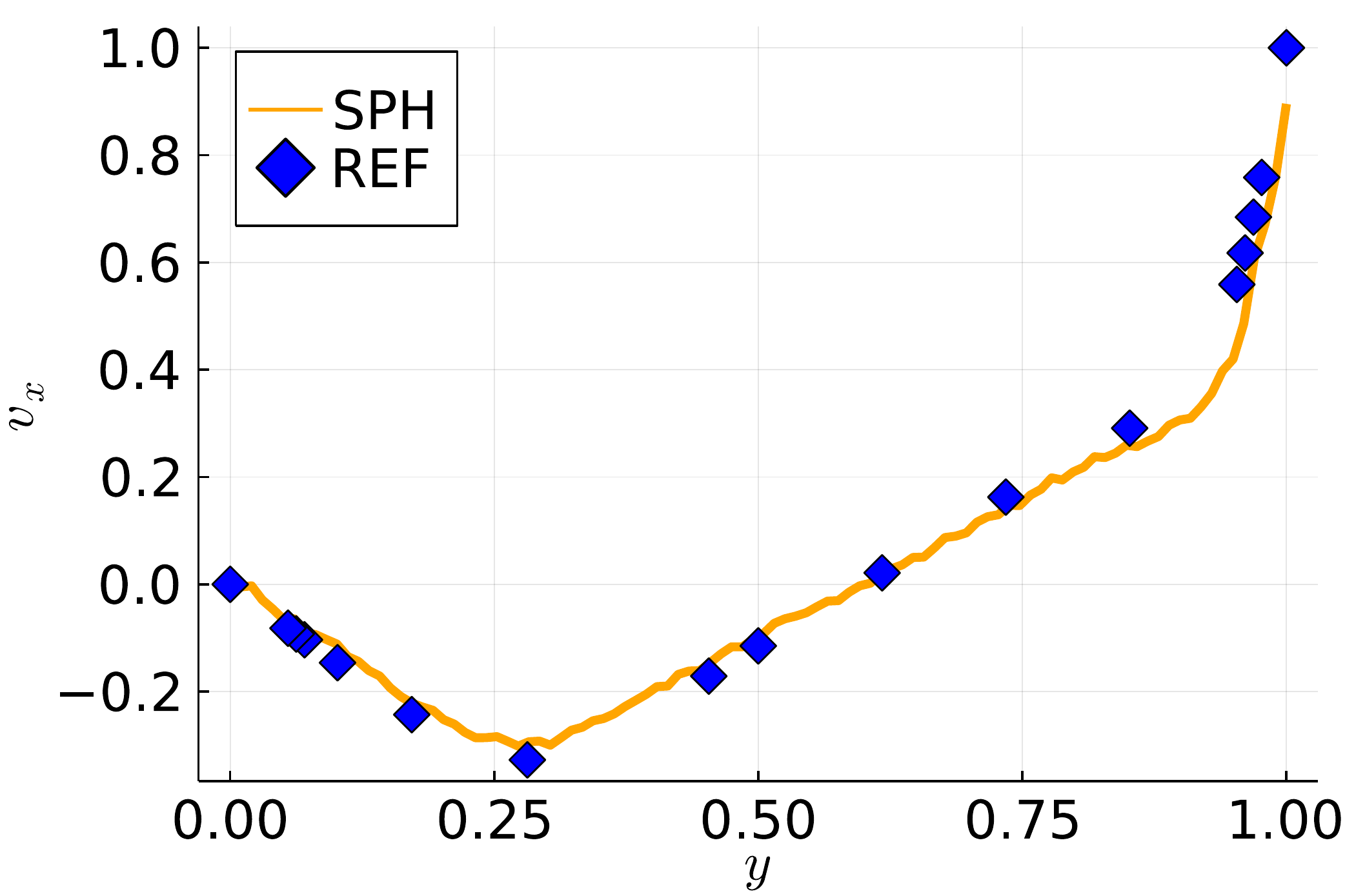}
		\caption{$v_x$ along $y$-centerline, $\mathrm{Re} = 400$, $t = 10$}
	\end{subfigure}
	\caption{Plot of transverse velocities along lines $x = \frac{1}{2}$ and $y = \frac{1}{2}$ in 
	the Newtonian lid-drive cavity benchmark. }
	\label{fig:ldc-graph}
\end{figure}
The advantages of the Taylor-Couette benchmark are simple implementation and availability of an 
exact solution (in steady state at least). It is, however, insufficient in the sense that the 
velocity field does not depend on the magnitude of $\nu$. For a more qualitative and challenging 
test, we include the lid-driven cavity benchmark. The geometry of this problem consists of a square
$ \Omega = (0,1) \times (0,1) $
filled with a viscous fluid. The left, right, and bottom boundaries are the walls with a no slip 
boundary condition, i.e. $\bm{v} = 0$, and the top boundary is moving at the prescribed velocity
$$ \bm{v}_\mathrm{lid} = \begin{pmatrix}
	1 \\ 0
\end{pmatrix}.$$  

For the shear energy, we use the constitutive equation \eqref{eq:DPR} with $\rho = 1$. Again, in 
the case of fluid flows, and especially incompressible flows, the 
ideal values for 
the shear and bulk speed of sound would be 
$c_s \gg 1 $,  $ c_0 \gg 1$
corresponding to the incompressible Navier-Stokes limit of the SHTC equations, i.e. $ \tau \sim  
\nu /c_s^2 \ll 1 $ and $ \mathrm{Ma} \sim \Vert\vv\Vert/c_0 \ll 1 $, but 
this is not possible in our scheme because the underlying ODE system \eqref{eq:odes-brief-relax} 
would become extremely stiff. Therefore, as an approximation, we set\footnote{We note that the 
shear sound speed $ c_s $ is not an artificial fitting parameter in the SHTC equations but it can 
be 
measured for real fluids via the sound dispersion data and fitted via a procedure described in 
\cite{HYP2016}.}
$$c_s = c_0 = 20.$$
We consider the cases of $\mathrm{Re} = 100$ and $\mathrm{Re} = 400$. The viscosity and the relaxation time are related to this number by:
$$ \nu = \frac{1}{\mathrm{Re}}, \qquad \tau = \frac{6 \nu}{ c_s^2}.$$  
The no slip walls are implemented as $h$-deep layer of particles with zero velocity. The lid is 
implemented similarly with immobile particles but ``pretending'' to have velocity 
$\bm{v}_\mathrm{lid}$ for the purposes of $\bbL$ computation (equation \eqref{eq:L}). The initial 
state is somewhat problematic in a weakly compressible scheme because discontinuities in the 
velocity field will generate shock waves. For this reason, we fix the zero lid velocity at $t = 0$ 
and gradually accelerate it up to 1.

The results are shown in Figures \ref{fig:ldc-v}, \ref{fig:ldc-A}. In Figure \ref{fig:ldc-graph}, 
we plot the transverse velocity along the center lines and compare the result to a referential 
solution \cite{ghia1982high}. Despite one can see slight discrepancies between SPH-SHTC solution 
and 
the reference one in Figure\,\ref{fig:ldc-graph} (which we suspect is caused by the problematic 
implementation of the Dirichlet 
boundary condition) a reasonable agreement between the solutions has 
been achieved.

\section{Conclusion}

We have developed a new SHTC-SPH numerical method that is suitable for simulations of both fluids 
and solids within a single framework. To the best of our knowledge it is the first ever 
discretization of the SHTC equations with an SPH scheme.   
The method discretizes (both in space and time) the Symmetric Hyperbolic Thermodynamically Compatible equations \eqref{eqn.PDE}, which describe both fluids and solids. First, we discretize them in space, which results in the SHTC-SPH ordinary differential equations \eqref{eq:odes}, which contain an evolution equation for a discrete analogue of the distortion field. Then, we prescribe a time integrator \eqref{eq:time-step}, which gives the SHTC-SPH numerical scheme. 

The scheme is then tested on benchmarks like a vibrating  Beryllium plate, twisting column, laminar 
Taylor-Couette flow, and lid-driven cavity flow, and shows acceptable agreement with the data, 
although finite volume and discontinuous Galerking ADER schemes \cite{DPRZ2016,LGPR2022} usually 
offer better precision.

In the future, we would like to investigate deeper geometrical properties of the SHTC-SPH scheme, 
such as its Hamiltonianity and conservation of Casimirs \cite{bb} as well as incorporating the heat 
conduction part of the SHTC equations. In addition, we plan to combine SHTC with the implicit 
variant of SPH for improved stability and performance.

\section*{Acknowledgments}
The authors are grateful to Markus Huetter for bringing our attention to the paper by Falk and 
Langer which helped us to construct the proposed discretization of the distortion field.
OK was supported by project No. START/SCI/053 of Charles University Research program.
MP was supported by project No. UNCE/SCI/023 of Charles University Research program. OK, MP and VK 
were also supported by the Czech Science Foundation (project no. 20-22092S).
IP is a member of the Gruppo Nazionale per il Calcolo
Scientifico of the Istituto Nazionale di Alta Matematica (INdAM GNCS) and acknowledges the 
financial support received from the Italian Ministry of Education, University and Research (MIUR) 
in the frame of the Departments of Excellence Initiative 2018–2022 attributed to the Department of 
Civil, Environmental and Mechanical Engineering (DICAM) of the University of Trento (Grant No. 
L.232/2016) and in the frame of the Progetti di Rilevante Interesse Nazionale (PRIN) 2017, Project 
No. 2017KKJP4X, “Innovative numerical methods for evolutionary partial differential equations and 
applications”.
\bibliographystyle{cas-model2-names}

\bibliography{library}

\end{document}